\documentclass[final]{elsarticle}

\usepackage{hyperref}
\usepackage{amsmath}
\usepackage{cases}

\newcommand{\PETSc}{\textsc{PETSc}}
\newcommand{\SUNDIALS}{\textsc{SUNDIALS}}

\usepackage{xcolor}
\usepackage{soul}
\definecolor{redhl}{rgb}{1,0.5,0.5}
\definecolor{bluehl}{rgb}{0.75,0.75,1}
\definecolor{greenhl}{rgb}{0.5,1.0,0.5}

\usepackage{subcaption}
\usepackage{placeins} 

\journal{Phys. Earth Planet. Inter.}

\biboptions{authoryear}

\begin{document}

\begin{frontmatter}

\title{Numerical solution of a non-linear conservation law applicable to the interior dynamics of partially molten planets}



\author[csh,eth]{Dan J. Bower\corref{mycorrespondingauthor}}
\ead{daniel.bower@csh.unibe.ch}
\author[eth,usi]{Patrick Sanan}
\ead{patrick.sanan@erdw.ethz.ch}
\author[uom]{Aaron S. Wolf}
\ead{aswolf@umich.edu}


\cortext[mycorrespondingauthor]{Corresponding author}

\address[csh]{Center for Space and Habitability, Uni. of Bern, Gesellschaftsstrasse 6, 3012 Bern, CH}
\address[eth]{Institute of Geophysics, ETH Z\"{u}rich, Sonneggstrasse 5, 8092 Z\"{u}rich, CH}
\address[uom]{Earth and Environmental Sciences, University of Michigan, 1100 North University Avenue, Ann Arbor, MI 48109-1005, USA}
\address[usi]{Institute of Computational Sciences, Universit\`{a} della Svizzera italiana, Via Giuseppe Buffi 13, 6900 Lugano, CH}


\begin{abstract}
The energy balance of a partially molten rocky planet can be expressed as a non-linear diffusion equation using mixing length theory to quantify heat transport by both convection and mixing of the melt and solid phases.  Crucially, in this formulation the effective or eddy diffusivity depends on the entropy gradient, $\partial S/\partial r$, as well as entropy itself.  First we present a simplified model with semi-analytical solutions that highlights the large dynamic range of $\partial S/\partial r$---around 12 orders of magnitude---for physically-relevant parameters.  It also elucidates the thermal structure of a magma ocean during the earliest stage of crystal formation.  This motivates the development of a simple yet stable numerical scheme able to capture the large dynamic range of $\partial S/\partial r$ and hence provide a flexible and robust method for time-integrating the energy equation.

Using insight gained from the simplified model, we consider a full model, which includes energy fluxes associated with convection, mixing, gravitational separation, and conduction that all depend on the thermophysical properties of the melt and solid phases.  This model is discretised and evolved by applying the finite volume method (FVM), allowing for extended precision calculations and using $\partial S/\partial r$ as the solution variable.  The FVM is well-suited to this problem since it is naturally energy conserving, flexible, and intuitive to incorporate arbitrary non-linear fluxes that rely on lookup data.  Special attention is given to the numerically challenging scenario in which crystals first form in the centre of a magma ocean.

The computational framework we devise is immediately applicable to modelling high melt fraction phenomena in Earth and planetary science research.  Furthermore, it provides a template for solving similar non-linear diffusion equations that arise in other science and engineering disciplines, particularly for non-linear functional forms of the diffusion coefficient.
\end{abstract}

\begin{keyword}
magma ocean \sep non-linear diffusion \sep finite volume method \sep mantle dynamics \sep mixing length theory
\end{keyword}


\end{frontmatter}


\section{Introduction}
Modelling diffusion has broad applications in science and engineering, but there is a tendency for the published literature to focus on diffusive systems that either possess convenient mathematical properties or are weakly non-linear.  Here, we are motivated by a physically-relevant and peculiar equation, similar to a non-linear diffusion equation, which arises from consideration of energy transport in a planetary interior.  Our solution method for this equation has immediate applicability to Earth and planetary sciences research for developing multi-scale and multi-physics models of the interior dynamics of planets.  Furthermore, the technique that we employ provides general insights for solving non-linear conservation laws and diffusion equations that extend to other disciplines.

Mixing length theory (MLT) enables energy transport by convection to be represented as a diffusive process, and this approach has been applied extensively to model the physics that governs the structure and evolution of stars \citep[e.g.,][]{KWW12}.  The theory estimates the contribution of eddy motions to bulk energy transfer by considering the distance that a parcel of material travels before it thermally equilibrates with its surroundings.  Often it is convenient to encapsulate this energy transfer within an ``effective'' or ``eddy'' diffusivity.  MLT was later applied to interrogate the cooling of an initially molten planet \citep{ABE93}.  As a molten planet cools from liquid to solid its viscosity increases by around 19 orders of magnitude and hence its dynamics transition from inviscid flow (liquid) to viscous creep (solid).  To model this process the eddy diffusivity switches between an inviscid and viscous flow scaling law based on the local Reynolds number.

Large astronomical objects such as stars and planets are close to global hydrostatic equilibrium and therefore 1-D radial models are appropriate for capturing the first-order structure and evolution of the interior of these bodies.  Modelling the radial structure is often the most scientifically informative approach, particularly when astrophysical data constraints are limited or model parameters poorly constrained.  Using MLT, 1-D models can replicate the general results of 3-D models \citep[e.g.,][]{KNK15} with significantly less calculation cost, thus permitting an extensive search of the parameter space.  An MLT scheme allows physical properties to be determined locally which enables properties to vary substantially throughout the 1-D domain.  In short, MLT remains a heavily utilised method for understanding planetary and stellar structure and also has applications in other fields.

The discovery of planets beyond the Solar System, so-called exoplanets, has thrust planet formation and evolution modelling to the forefront of planetary science.  In particular, a large fraction of detected exoplanets are rocky (i.e., akin to Earth) and may have evolved from an initially molten state due to the heat generated by accretion, core formation, short-lived radioisotopes, and late-stage impacts \citep[e.g.,][]{S14}.  Furthermore, some exoplanets are sufficiently close to their star that they retain a permanent magma ocean on their day-side.  Hence, developing techniques and methods to model high melt fraction phenomena is crucial to advance modelling capabilities and enable us to bridge the dynamic timescale between melt (inviscid) and solid (viscous) processes.  In particular, rocky exoplanets demand a flexible modelling strategy because their size---in terms of mass, radius, and internal pressure---can far exceed those of the terrestrial planets in the Solar System, and their composition may be strikingly different from Earth \citep[e.g.,][]{MLM12}.

Incorporating high melt fraction dynamics in existing 2-D geodynamic viscous flow codes is a recent development \citep[e.g.,][]{TLF17}.  For simplicity and ease, the dynamics are typically encapsulated using a constant eddy diffusivity to model the enhanced energy transport due to the advection of melt \citep[e.g.,][]{NBS14}.  Other energy transport mechanisms that operate during the high melt fraction regime, such as convective mixing, are often excluded.  These simplifications are often a necessary compromise to limit the computational expense for models that are already teeming with physical and chemical complexity.  Similarly, resolving the ultra-thin thermal boundary layer at the top of a magma ocean ($\sim$ few cms) \citep[e.g.,][]{ABE93} is beyond the computational capabilities of 2-D models.  Interestingly, even studies that use quasi 1-D dynamic interior models invoke similar simplifications, often because they are primarily concerned with the evolution of the atmosphere \citep[e.g.,][]{HAG13,LMC13,HKA15}.

Therefore, we are motivated to revisit the MLT formulation for a partially molten planet \citep{ABE93} and analyse the full functional form (without simplification) of the eddy diffusivity.  The eddy diffusivity not only depends on material properties, which themselves are a function of the local thermodynamic state, but also on the local super-adiabatic temperature gradient or equivalently local entropy gradient.  This gradient-dependence to the diffusivity introduces numerical precision challenges that are typically absent from simpler diffusive process models.  The insights that we gain about the behaviour of the system enable us to devise a fast and flexible code that is amenable to large-scale simulation of high melt fraction dynamics in planetary bodies.  The code can be used to explore parameter space, test modelling intuition, and compare with results from 2-D simulations.  It further provides techniques for improving the stability, speed and precision of high melt fraction dynamics in high dimension (2-D+) simulations.
\section{Physical model}
\subsection{Fundamental equations}
\label{sect:fundamental}
The basic equations for modelling energy transport in a planetary interior are provided by \cite{ABE93,ABE95,ABE97}.  These chart the evolution of a planet that is initially fully molten and during subsequent cooling evolves to a partially molten state where melt and solid crystals coexist.  Here, we express the non-linear conservation of thermal energy in integral form in terms of entropy:
\begin{equation}
\label{eqn:entropy_integral}
\int_V \rho T \dfrac{\partial S}{\partial t} dV = -\int_A F \cdot n dA + \int_V \rho H dV
\end{equation}
where $S$ is specific entropy, $\rho$ density, $T$ temperature, $F$ heat flux vector, $n$ outward unit normal to the bounding surface(s) $A$, $H$ internal heat generation per unit mass, $t$ time, and $V$ volume.  We formulate the problem using entropy because it is a natural coordinate for both convecting systems, which are near-adiabatic, and thermodynamic models of mantle melting \citep[see][for discussion]{SA07}.  Total heat flux is:
\begin{equation}
\label{eqn:heat_flux}
F = F_{\rm conv} + F_{\rm mix} + F_{\rm cond} + F_{\rm grav}
\end{equation}
where $F_{\rm conv}$ is the convective flux, $F_{\rm mix}$ is the flux due to the mixing of melt and solid, $F_{\rm cond}$ is the conductive flux, and $F_{\rm grav}$ is the flux due to gravitational separation of melt and solid.  For a 1-D system where radius $r$ is the spatial coordinate:
\begin{equation}
\label{eqn:Fconv}
F_{\rm conv} = -\rho T \kappa_h \frac{\partial S}{\partial r}
\end{equation}
where the eddy diffusivity $\kappa_h$ is estimated using the average mean free path of convective parcels:
\begin{equation}
\kappa_h \sim u l
\end{equation}
where $u$ is a characteristic velocity that is derived from force-balance considerations and mixing length theory, and $l$ is the mixing length.  We choose the mixing length to be the distance from the nearest thermal boundary layer so that the calculated heat flux fits experimental results \cite[Fig. 3 in][]{ABE95, SC97}.  Therefore, in the simplest case of single layer convection, the mixing length at a given radius is the minimum distance from the top or bottom surface.  Post-magma ocean crystallisation, viscous creep is the norm for planetary convection where the rise and fall of convective parcels is primarily resisted by viscous drag.  Therefore, the Stokes settling velocity of convective parcels is \citep[e.g.,][]{SN86}:
\begin{subnumcases}{u_{\rm visc}=\label{eqn:uvisc}}
\frac{\alpha g l^3 T}{18 \nu c} \dfrac{\partial S}{\partial r} & \text{for } $\partial S/\partial r<0$ \label{eqn:uvisc_conv}\\
0 & \text{otherwise} \label{eqn:uvisc_none}
\end{subnumcases}
where $\alpha$ is the thermal expansion coefficient, $c$ specific heat capacity, $g$ gravity (negative by convention), and $\nu=\eta/\rho$ kinematic viscosity.  The dynamic viscosity $\eta$ of the aggregate is given by Eq.~\ref{eqn:visc_nonskew}.  Here, Eq.~\ref{eqn:uvisc_none} stipulates that the adiabatic temperature gradient must be exceeded for convection to occur.  The Reynolds number is:
\begin{equation}
Re = u_{\rm visc} l / \nu
\end{equation}
In highly turbulent systems, such as a fully molten magma ocean, dynamic pressure rather than viscous drag is the dominant force resisting the vertical transport of convective parcels.  This gives rise to a characteristic inviscid flow velocity \citep{V53}:
\begin{subnumcases}{u_{\rm invis}=\label{eqn:uinvis}}
\sqrt{\frac{\alpha g l^2 T}{16 c}\dfrac{\partial S}{\partial r}} & \text{for } $\partial S/\partial r<0$ \label{eqn:uinvis_conv}\\
0 & \text{otherwise} \label{eqn:uinvis_none}
\end{subnumcases}
By defining a critical Reynolds number $Re_{\rm crit}=9/8$ \citep[e.g.,][]{ABE95} we can construct a piecewise function for $\kappa_h$ that switches between the viscous and inviscid velocity scaling depending on the local Reynolds number\footnote{We prefer this representation of $\kappa_h$ using $Re$ due to its physical insight but the formulation is identical to Eq.~15 in \cite{ABE93} and Eq.~6 in \cite{ABE97}.  Eq.~47 in \cite{ABE95} also gives the same $\kappa_h$, although Eq.~47c erroneously has $\rho C_p$ inside the square root.}.  Note the strong non-linear sensitivity of $\kappa_h$ to $l$; for $Re_{\rm crit}<Re$ it scales as $l^2$ and otherwise as $l^4$:
\begin{subnumcases}{\label{eqn:kappah} \kappa_h=}
    u_{\rm invis}l & \text{for } $Re_{\rm crit} < Re$  \label{eqn:kappah_inviscid} \\
    u_{\rm visc}l & \text{for }  $0 \le Re \le Re_{\rm crit}$  \label{eqn:kappah_viscous} \\
    0 & \text{for } $Re=0$
\end{subnumcases}
Convective mixing is described using Fick's law and quantifies latent heat transport as crystals form and remelt as they are displaced quasi-adiabatically by convective flow\footnote{Eq.~9 in \cite{ABE97} should include $\rho$ as in Eq.~14 in \cite{ABE93} and Eq.~56 in \cite{ABE95}}:
\begin{equation}
\label{eqn:Fmix}
F_{\rm mix} = -\rho T \Delta S_{\rm fus} \kappa_h \dfrac{\partial \phi}{\partial r}
\end{equation}
where $\Delta S_{\rm fus}=S_{\rm liq}- S_{\rm sol}$ is the entropy of fusion and $S_{\rm liq}$ and $S_{\rm sol}$ are the liquidus and solidus, respectively.  Melt fraction $\phi$ is:
\begin{subnumcases}{\label{eqn:meltfraction} \phi=}
1 & \text{for } $S_{\rm liq}<S$\\
 \frac{S - S_{\rm sol}}{\Delta S_{\rm fus}} & \text{for} $S_{\rm sol} \le S \le S_{\rm liq}$ \label{eqn:g_meltfraction}\\
0 & \text{for } $S<S_{\rm sol}$
\end{subnumcases}
Conduction is determined by Fourier's law:
\begin{equation}
\label{eqn:Fcond}
F_{\rm cond} =  - \rho T \kappa \left( \frac{\partial S}{\partial r} \right)-\rho c \kappa \left( \frac{\partial T}{\partial r} \right)_S
\end{equation}
where $\kappa$ is the thermal diffusivity and $(\partial T/\partial r)_S$ is the adiabatic temperature gradient.  Gravitational separation occurs by permeable flow of melt in the solid matrix at small melt fraction and crystal settling or floatation at large melt fraction:
\begin{equation}
\label{eqn:Fgrav}
F_{\rm grav} = \frac{a^2 g \rho (\rho_{\rm liq}-\rho_{\rm sol})\zeta_{\rm grav}(\phi)}{\eta_m} T \Delta S_{\rm fus}
\end{equation}
where $a$ is the grain size, $\eta_m$ melt viscosity, and subscripts ``liq'' and ``sol'' denote that the quantity is evaluated at the liquidus and solidus, respectively.  The flow mechanism factor $\zeta_{\rm grav}$ depends on the flow law and is a function of melt fraction\footnote{$\zeta_{\rm grav}$ is the same as $F(\phi)$ given in Eq.~46 in \cite{ABE95} and Eq.~8 in \cite{ABE97}.  However, with this definition for $F(\phi)$, both Eq.~45 in \cite{ABE95} and Eq.~7 in \cite{ABE97} disagree with Eq.~12 in \cite{ABE93} and furthermore do not have the correct units.  We therefore rederive the expressions following the outline in \cite{ABE95} and find agreement with \cite{ABE93}.  We suspect that both Eq.~45 in \cite{ABE95} and Eq.~7 in \cite{ABE97} were not updated to account for the different formulation of $F(\phi)$ compared to that in \cite{ABE93}.  This is implied by the remark in \cite{ABE95} that ``the definition of $F(\phi)$ is different''.}:
\begin{subnumcases}{\zeta_{\rm grav}=\label{eqn:F_melt_fraction}}
    \dfrac{2}{9} \phi (1-\phi) & $\text{for } \dfrac{\rho_{\rm liq}}{0.29624\rho_{\rm sol}+\rho_{\rm liq}} \le \phi$ \label{eqn:grav_stokes}\\
     \dfrac{5}{7} \dfrac{{\rho_{\rm sol}}^{9/2} \phi^{11/2}(1-\phi)}{(\rho_{\rm liq}+(\rho_{\rm sol}-\rho_{\rm liq})\phi )^{9/2}} & $\text{for } \dfrac{\rho_{\rm liq}}{11.993\rho_{\rm sol}+\rho_{\rm liq}} \le \phi < \dfrac{\rho_{\rm liq}}{0.29624\rho_{\rm sol}+\rho_{\rm liq}}$ \label{eqn:grav_rumpf-gupte}\\
    \dfrac{1}{1000} \dfrac{{\rho_{\rm sol}}^2 \phi^3}{{\rho_{\rm liq}}^2(1-\phi)} & $\text{for } \phi < \dfrac{\rho_{\rm liq}}{11.993\rho_{\rm sol}+\rho_{\rm liq}}$ \label{eqn:grav_blake-kozeny-carman}
\end{subnumcases}
Eq.~\ref{eqn:grav_stokes} is derived by considering the Stokes' velocity for spherical crystals.  Eqs.~\ref{eqn:grav_rumpf-gupte} and ~\ref{eqn:grav_blake-kozeny-carman} arise from permeability flow laws given by Rumpf-Gupte \citep{RG71} and Blake-Kozeny-Carman, respectively \citep[see][for discussion]{ABE95}.  From Eq.~\ref{eqn:F_melt_fraction} it is clear that gravitational separation only occurs in the mixed phase region, as expected, since $F_{\rm grav}=0$ for $\phi=0$ or $\phi=1$.  If melt is less dense than solid ($\rho_{\rm liq}<\rho_{\rm sol}$) then $F_{\rm grav}$ is positive and heat is transported upwards toward the top surface.  But if a melt-solid density crossover exists in the mantle then $F_{\rm grav}$ can be negative for certain radii (depths) and heat is carried down towards the core-mantle boundary.

The formulation for the dynamic viscosity is designed to capture the rheological transition where the aggregate viscosity changes fairly abruptly between the melt and solid viscosity at a critical melt fraction.  Our formulation captures the trend observed in the semi-empirical model of \cite{CCB09}, although we use the end-member melt and solid viscosities in \cite{ABE93}.  The viscosity of the aggregate $\eta$ is:
\begin{equation}
\label{eqn:visc_nonskew}
\log_{10}{ \eta} = z \log_{10}( \eta_m) + (1-z) \log_{10}( \eta_s)
\end{equation}
where $\eta_m$ and $\eta_s$ are the melt and solid viscosity, respectively, and $z$ is the transition function:
\begin{subequations}
\begin{alignat}{2}
z(y)=& \dfrac{1}{2} \left(1+ \tanh(y) \right) \label{eqn:tanh}\\
y(\phi) =& \frac{\phi-\phi_c}{\phi_w} \label{eqn:tanh_arg}
\end{alignat}
\label{eqn:smoothing}
\end{subequations}
where $\phi_c$ is the critical melt fraction and $\phi_w$ is the transition width.  We choose the planetary radius $R_0$ and a reference entropy $S_0$, temperature $T_0$, and density $\rho_0$, to non-dimensionalise the equations.  For convenience we choose reference values that correspond to the maximum of the liquidus in \cite{SKS09} where $dS_{\rm liq}/dr=0$.  The primary scalings are given in Table~\ref{tbl:nondim} and others are straightforward to derive.

\begin{table}[t]
\centering
\begin{tabular}{| l | l | l | l |}
\hline
Parameter & Scaling & Value & Units \\
\hline
Acceleration & $S_0 T_0 {R_0}^{-1}$ & 1.894 & m s$^{-2}$\\
Area & ${R_0}^2$ & $4.059 \times 10^{13}$ & m$^{2}$\\
Conductivity & $S_0 \rho_0 R_0 (S_0 T_0)^{-1/2}$ & $3.056 \times 10^{17}$ & W m$^{-1}$ K$^{-1}$\\
Density & $\rho_0$ & 4613 & kg m$^{-3}$\\
Diffusivity & $R_0 (S_0 T_0)^{1/2}$ & $2.214 \times 10^{10}$ & m$^2$ s$^{-1}$\\
Energy & $S_0 \rho_0 T_0 {R_0}^3$ & $1.440 \times 10^{31}$ & J\\
Entropy & $S_0$ & 2993 & J kg$^{-1}$ K$^{-1}$\\
Flux & $\rho_0 (S_0 T_0)^{3/2}$ & $1.935 \times 10^{14}$ & W m$^{-2}$\\
Heat generation & $(S_0 T_0)^{3/2} {R_0}^{-1}$ & 6584 & W kg$^{-1}$\\ 
Mass & $\rho_0 {R_0}^3$ & $1.193 \times 10^{24}$ & kg\\
Pressure & $\rho_0 S_0 T_0$ & $5.569 \times 10^{10}$ & Pa\\
Radius & $R_0$ & $6.371 \times 10^{6}$ & m\\
Temperature & $T_0$ & 4034 & K\\
Time & $R_0 (S_0 T_0)^{-1/2}$ & 1834 & s\\
Viscosity & $\rho_0 R_0 (S_0 T_0)^{-1/2}$ & $1.021 \times 10^{14}$ & Pa s\\
Volume & ${R_0}^3$ & $2.586 \times 10^{20}$ & m$^{3}$\\
\hline
\end{tabular}
\caption{Dimensional scalings of primary parameters.}
\label{tbl:nondim}
\end{table}

\subsection{Pressure and material properties}
\label{sect:material}
The fundamental equations (Section~\ref{sect:fundamental}) are expressed in terms of radius $r$ and it is necessary to relate this to hydrostatic pressure $P$ to interface the evolution with equations of state for planetary materials.  We determine the hydrostatic pressure within a planet as a function of depth using an equation of state (EOS).  An appropriate choice is the Adams-Williamson EOS, which assumes adiabatic compression of a chemically homogenous material:
\begin{equation}
\label{eqn:hydro_pressure}
P = -\frac{\rho_r g}{\beta} ( e^{{\beta z}}-1)
\end{equation}
where $\rho_r$ is a reference surface density, $\beta$ is a measure of the compressibility of the material, and $z$ is depth.  For Earth, gravity is near constant throughout the mantle and we solve for $\rho_r$ and $\beta$ using a least-squares fit to the density profile of the lower mantle \citep{DA81}.

We use a thermodynamic description for a mantle composed of MgSiO$_3$, which can exist as a melt ($S>S_{\rm liq}$), solid ($S<S_{\rm sol}$), or partially molten aggregate ($S_{\rm sol} \le S \le S_{\rm liq}$).  Solid and melt thermophysical properties are determined by \cite{M09} and \cite{WB17}, respectively.  The melt--solid density contrast is on average -5\% in the lower mantle and since we only consider a single component there is no density crossover and the melt is less dense than the solid everywhere.  Order-of-magnitude estimates for the thermophysical properties of the aggregate are derived by considering an ideal solution that assumes linear additivity \citep[e.g.,][]{SOLO07}.  It is advantageous to pre-calculate lookup tables of material properties that can subsequently be queried during the evolution of a model, although analytical expressions could also be used if available.  This ensures the model retains flexibility to incorporate any dataset of material properties and eliminates computational overhead associated with Gibb's free energy minimisation that would otherwise be required to compute chemical and phase equilibria.  Therefore, from the perspective of the numerical scheme that we subsequently develop (Section~\ref{sect:numerical}), we treat thermodynamic quantities (including temperature) as lookup quantities that are a function of entropy $S$ and pressure $P$.

For the full model (Section~\ref{sect:results}) we use lookup tables for melt and solid properties with an approximate resolution of 23 Jkg$^{-1}$K$^{-1}$ in entropy and 2 GPa in pressure.  The resolution is chosen according to the smoothness of the data, but since it is difficult to predict the influence of the input data in a non-linear model we run three additional cases with coarser lookup tables and compare the output (see supplementary material).  In fact, thermodynamic considerations may result in discontinuous material properties across $S_{\rm liq}$ and $S_{\rm sol}$ and this is typically the source of the largest gradients in material properties.  For this reason we use the transition function (Eq.~\ref{eqn:tanh}) to ensure properties vary smoothly across the liquidus and solidus to avoid numerical difficulties.  It is further used to smoothly introduce the convective mixing flux $F_{\rm mix}$ as the system cools below the liquidus and enters the mixed phase region.  The smoothing is formulated akin to the viscosity transition (Eq.~\ref{eqn:smoothing}) where now $\phi$ is the generalised melt fraction (Eq.~\ref{eqn:g_meltfraction}), $\phi_c$ is 1 and 0 for smoothing across the liquidus and solidus, respectively, and the smoothing width $\phi_w=0.01$ which corresponds to $\approx$ 8 Jkg$^{-1}$K$^{-1}$ in the lower mantle.  See supplementary material for discussion on the sensitivity of our results to the smoothing width.
\subsection{Boundary conditions}
\label{sect:bc}
Our objective is to devise a flexible framework to probe a variety of planetary cooling scenarios and therefore it is useful to consider both linear and non-linear boundary conditions.  For the top surface, an isothermal boundary condition is appropriate for a planet in radiative equilibrium where the surface temperature is dictated by incoming solar radiation rather than interior heat.  Another simple choice is a constant heat flux boundary condition.  However, for the earliest stage of planetary cooling the surface radiates as a grey-body \citep[e.g.,][]{ET08}:
\begin{equation}
\label{eqn:bc_radiative}
F_{\rm surf} = \epsilon \sigma ({T_{\rm surf}}^4-{T_{\rm eq}}^4)
\end{equation}
where $F_{\rm surf}$ and $T_{\rm surf}$ are surface heat flux and temperature, respectively, $\epsilon$ the emissivity which is unity in the absence of atmospheric effects, and $\sigma$ the Stefan-Boltzmann constant.  $T_{\rm eq}$ is the radiative equilibrium temperature of the planet, which is approximately 273 K for Earth.  Both mixing length and boundary layer theory predict that a magma ocean has an ultra-thin thermal boundary layer at the top surface with a thickness of a few centimetres.  The temperature drop across the boundary layer $\Delta T_{\rm bl}$ is parameterised \citep[e.g.,][]{VSS00,RS06}:
\begin{equation}
\label{eqn:bc_utbl}
\Delta T_{\rm bl} = b {T_{\rm surf}}^3
\end{equation}
For the bottom surface, which corresponds to the core-mantle boundary (CMB), we consider the energy balance for the core, neglecting the energy contribution from growth of the inner core and internal heat sources:
\begin{equation}
\label{eqn:core_energy}
m_{\rm core} c_{\rm core} \dfrac{dT_{\rm core}}{dt} = -4 \pi r_{\rm cmb}^2 F_{\rm cmb}
\end{equation}
where $m_{\rm core}$ and $c_{\rm core}$ are the mass and heat capacity of the core, respectively, $T_{\rm core}$ is the mass-weighted effective temperature of the core, $r_{\rm cmb}$ is the planetary radius at the CMB, and $F_{\rm cmb}$ is the CMB heat flux.  Note that the core and mantle are thermally coupled which is why we must consider temperature rather than entropy in the formulation of this boundary condition.  It is convenient to relate the bulk temperature of the core to the temperature at the CMB:
\begin{equation}
\label{eqn:Tcore}
T_{\rm core} = \hat{T}_{\rm core} T_{\rm cmb}
\end{equation}
where $\hat{T}_{\rm core}$ is a constant that accounts for the thermal profile of the core, and $T_{\rm cmb}$ is the CMB temperature, which formally is the foot temperature of the core adiabat.  We derive $\hat{T}_{\rm core}$ by assuming the core is isentropic (i.e., vigorously convecting) with Gr\"{u}neisen parameter $\gamma=1.3$ \citep{STACEY94,LPL01}.  Changes in the mass distribution of the core due to cooling are negligible compared to changes in temperature.  Therefore, the core density profile (time-independent) is given by the Preliminary Reference Earth Model \citep{DA81} which we modify to exclude the compositional variation of the inner core.  These considerations result in a thermal structure correction factor of $\hat{T}_{\rm core}=1.147$ and the boundary condition becomes:
\begin{equation}
\label{eqn:bc_cmb}
\dfrac{dT_{\rm cmb}}{dt} = -\dfrac{4\pi r_{\rm cmb}^2 F_{\rm cmb}}{m_{\rm core} c_{\rm core} \hat{T}_{\rm core}}
\end{equation}
\section{Simplified model}
\label{sect:simplified}
\subsection{Physical description}
We present a simplified model, derived from the fundamental equations (Section~\ref{sect:fundamental}), which captures key aspects of the full system and highlights the relationship between the non-linear conservation law and non-linear diffusion.  The insights gained from this analysis guide our selection of the numerical method that we implement to solve the full model.  The strong form of the conservation law (Eq.~\ref{eqn:entropy_integral}) is:
\begin{equation}
\rho T \frac{\partial S}{\partial t} = - \nabla \cdot F + \rho H
\end{equation}
Assuming that $F$ is a purely radial function (along with $\rho$, $T$) and excluding internal heat sources ($H=0$), implies that in spherical coordinates:
\begin{equation}
\label{eqn:spherical_cons}
\rho T \frac{\partial S}{\partial t} = - \frac{1}{r^2} \frac{\partial}{\partial r}r^2 F
\end{equation}
We neglect variations in $\rho T$ and assume that the domain of interest has sufficiently little variation in radius to ignore the spherical geometric terms such that the problem is locally 1-D:
\begin{equation}
\label{eqn:pm_cons}
\frac{\partial S}{\partial t} = - \frac{\partial F}{\partial r}
\end{equation}
We further restrict our interest to the centre of a partially molten mantle just below the liquidus and away from thermal boundary layers.  Here, gravitational separation is insignificant due to efficient mixing at high Rayleigh number and conduction is negligible in comparison to convection.  Therefore, the total flux is dominated by convection (Eq.~\ref{eqn:Fconv}) and mixing (Eq.~\ref{eqn:Fmix}).
\begin{subequations}
\begin{alignat}{2}
\widetilde{F}=& \widetilde{F}_{\rm conv} + \widetilde{F}_{\rm mix} \label{eqn:Ftotal_toy}\\
\widetilde{F}_{\rm conv}=& -\widetilde{\kappa}_h \frac{\partial S}{\partial r}  \label{eqn:Fconv_toy}\\
\widetilde{F}_{\rm mix}=&-\Delta S_{\rm fus} \widetilde{\kappa}_h \dfrac{\partial \phi}{dr} \approx -\widetilde{\kappa}_h \left( \frac{\partial S}{\partial r}-\frac{dS_{\rm liq}}{dr} \right) \label{eqn:Fmix_toy}
\end{alignat}
\label{eqn:Flux_toy}
\end{subequations}
In Eq.~\ref{eqn:Fmix_toy}, $\partial \phi/dr$ reduces to a simple expression of $\Delta S_{\rm fus}$, $\partial S/\partial r$ and $dS_{\rm liq}/dr$ using Eq.~\ref{eqn:meltfraction} and then we apply the approximation that $\phi=1$ near the liquidus.  Flow is inviscid close to the liquidus so the eddy diffusivity $\widetilde{\kappa}_h$ is:
\begin{subnumcases}{\widetilde{\kappa}_h=\frac{1}{64}\label{eqn:pm_kh}}
  \sqrt{-\frac{\partial S}{\partial r}} & \text{for } $\frac{\partial S}{\partial r}<0$ \label{eqn:pm_khconv} \\
  0 & \text{for } $\frac{\partial S}{\partial r} \ge 0$ \label{eqn:pm_khnone}
\end{subnumcases}
For Earth, the (non-dimensional) mixing length $l \approx 1/4$ since the core and mantle thickness are both approximately half of the planetary radius.  This gives the prefactor of $1/64$ when combined with the $1/16$ constant inside the square root in Eq.~\ref{eqn:uinvis_conv}.  The non-linear diffusion coefficient $\widetilde{\kappa}_h$ depends on $\partial S/\partial r$ via a square-root non-linearity (Eq.~\ref{eqn:pm_khconv}); for small $\partial S/\partial r$, this has the effect of strongly amplifying any error in $\partial S/\partial r$.  In the full set of equations it also depends on $S$ and $r$ through its dependence on material parameters (Eqs.~\ref{eqn:uvisc_conv}, \ref{eqn:uinvis_conv}).  Crucially, $\widetilde{\kappa}_h$ enforces a strong asymmetry because non-trivial solutions---defined by non-zero convective and mixing flux---are only admissible when $\partial S/\partial r<0$.
\subsection{Steady-state analysis}
Inspection of the flux (Eq.~\ref{eqn:Flux_toy}) reveals a solution space where $\widetilde{F}$ can either be positive (radially outward), negative (radially inward), or zero:
\begin{subequations}
\begin{alignat}{4}
\widetilde{F} > 0 & \quad \text{ for } \quad & \frac{\partial S}{\partial r}<\frac{1}{2}\frac{dS_{\rm liq}}{dr} & \quad \text{ and } \quad &\frac{\partial S}{\partial r}<0 \label{eqn:Fg0}\\
\widetilde{F} < 0 & \quad \text{ for } & \frac{\partial S}{\partial r}>\frac{1}{2}\frac{dS_{\rm liq}}{dr} & \quad \text{ and } \quad &\frac{\partial S}{\partial r}<0 \label{eqn:Fl0}\\
\widetilde{F} = 0 & \quad \text{ for } & \frac{\partial S}{\partial r}=\frac{1}{2}\frac{dS_{\rm liq}}{dr} & \quad \text{ or } \quad &\frac{\partial S}{\partial r} \ge 0 \label{eqn:Fe0}
\end{alignat}
\label{eqn:F_sign}
\end{subequations}
For $dS_{\rm liq}/dr<0$ both negative and positive fluxes are admissible since Eq.~\ref{eqn:Fg0} and \ref{eqn:Fl0} can both be satisfied for a different $\partial S/\partial r$.  For $dS_{\rm liq}/dr>0$, however, only a positive flux is permitted since Eq.~\ref{eqn:Fl0} can otherwise never be satisfied.  This asymmetry arises because $\widetilde{F}_{\rm conv}$ is always positive regardless of the sign of $dS_{\rm liq}/dr$, but the total flux $\widetilde{F}$ can be negative if mixing overwhelms convection, with $|\widetilde{F}_{\rm mix}|>|\widetilde{F}_{\rm conv}|$ and $dS_{\rm liq}/dr<0$.  In this situation, mixing enables heat to be buried deep in the interior by transfer of latent heat.  However, the capacity to transport energy downwards towards the CMB is restricted by the availability of latent heat.  In this simplified model, latent heat is manifest through the relative difference between the entropy profile ($\partial S/\partial r$) and the liquidus ($dS_{\rm liq}/dr)$.  We compute the derivative of $\widetilde{F}$ (Eq.~\ref{eqn:Ftotal_toy}) with respect to $\partial S/\partial r$ and determine a minimum (i.e., largest negative) flux:
\begin{equation}
\widetilde{F}_{\rm min} = \frac{\frac{dS_{\rm liq}}{dr} \sqrt{-\frac{dS_{\rm liq}}{dr}} }{96 \sqrt{6}} \quad \text{ where } \frac{\partial S}{\partial r}=\frac{1}{6} \frac{dS_{\rm liq}}{dr}
\label{eqn:Fmin_toy}
\end{equation}
In essence, a small negative entropy gradient can drive a net negative flux, but once the magnitude of $\partial S/\partial r$ becomes large then convective heat transport dominates and the net flux reverts to positive.  Note that there is no maximum (largest positive) flux because increasing the magnitude of $\partial S/\partial r$ can increase the (positive) flux without bound.

\begin{figure}[t]
\centering
\includegraphics[width=0.8\textwidth]{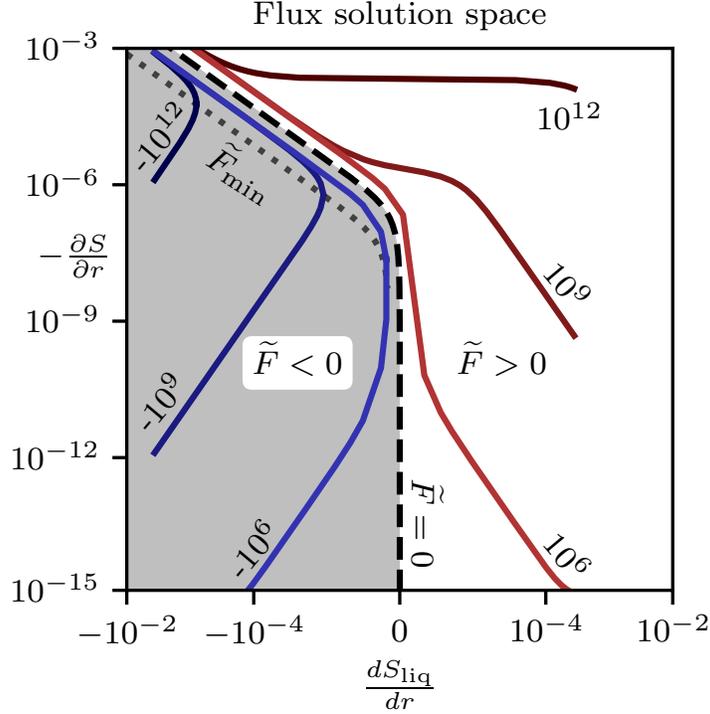}
\caption{Flux solution space for the simplified model.  The grey shaded region denotes $\widetilde{F}<0$ (Eq.~\ref{eqn:Fl0}), white is $\widetilde{F}>0$ (Eq.~\ref{eqn:Fg0}), black dashed line is $\widetilde{F}=0$ (Eq.~\ref{eqn:Fe0}), and grey dotted line is $\widetilde{F}=\widetilde{F}_{\rm min}$ (Eq.~\ref{eqn:Fmin_toy}).  Positive (10$^6$, 10$^9$, 10$^{12}$) and negative (-10$^6$, -10$^9$, -10$^{12}$) fluxes (Wm$^{-2}$) are offset from the asymptote (black dashed line) for visual clarity.  Axes are arcsinh transformed to capture both negative and positive values and the large dynamic range of the solutions.}
\label{fig:regime}
\end{figure}

We plot the solution space for $\widetilde{F}$ as a function of $dS_{\rm liq}/dr$ and $\partial S/\partial r$ and mark the regions defined by Eqs.~\ref{eqn:F_sign} and \ref{eqn:Fmin_toy} (Fig.~\ref{fig:regime}).  We then compute steady-state solutions to Eq.~\ref{eqn:pm_cons} ($\partial S/\partial t=0$) for constant positive and negative fluxes.  For $\widetilde{F}_{\rm min}<\widetilde{F}<0$ there are multiple solutions---for a given $dS_{\rm liq}/dr$ a negative flux can be accommodated by either a small or large negative $\partial S/\partial r$.  However, the physical relevance of $\widetilde{F}<0$ is questionable and this region can likely be excluded from further consideration for two reasons.  Firstly, a cooling molten planet radiates as a grey body and thus has a positive heat flux boundary condition at the surface.  Secondly, only $\widetilde{F}>0$ provides a globally continuous steady state solution for all $dS_{\rm liq}/dr$.

As previously mentioned, the simplified model does not inherently apply an upper limit to the (positive) flux that a magma ocean can transport.  This is evident in Fig.~\ref{fig:regime} where a large positive flux (10$^{12}$ Wm$^{-2}$) is accommodated by a comparatively large $\partial S/\partial r$.  For $dS_{\rm liq}/dr<0$, positive flux is always accommodated by $\partial S/\partial r$ that is less than half the gradient of the liquidus.  Crucially, this constrains the dynamic range of $\partial S/\partial r$ within about 4 orders of magnitude regardless of the magnitude of $\widetilde{F}$.  For $dS_{\rm liq}/dr>0$, however, latent heat transport is sufficiently large and positive that the system can only accommodate reduced fluxes by driving $\partial S/\partial r$ toward zero to minimise $\widetilde{\kappa}_h$ (Eq.~\ref{eqn:pm_kh}) and hence the total flux.  This is important because the total flux is therefore limited by the efficiency of radiative heat transfer to space.  An appropriate estimate is 10$^6$ Wm$^{-2}$, corresponding to a black body temperature of 2050 K which is compatible with the expected surface temperature of a magma ocean.  This flux predicts a large dynamic range of $\partial S/\partial r$ of about 12 orders of magnitude.

\begin{figure}[t]
\centering
\includegraphics[width=0.8\textwidth]{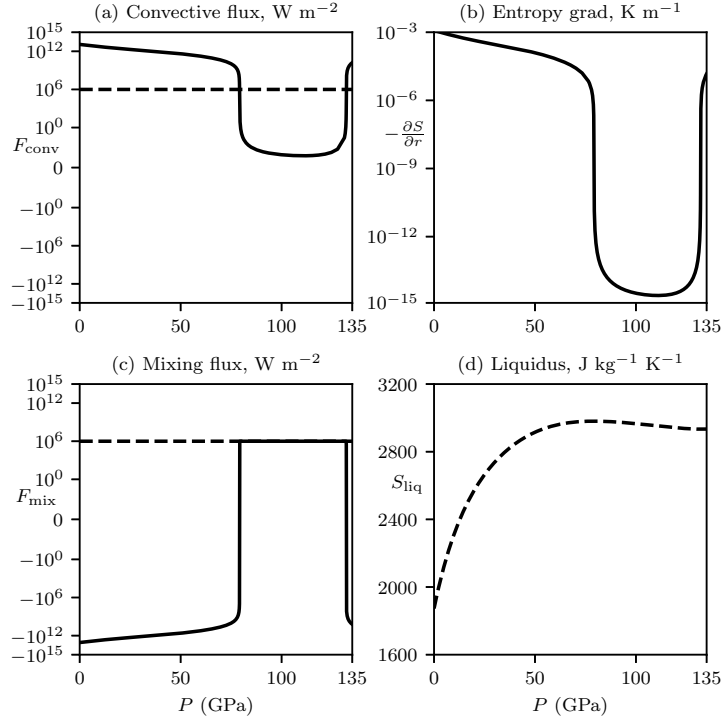}
\caption{Semi-analytical steady state solution to the simplified model using the liquidus for the middle-out crystallisation scenario from \cite{SKS09} and prescribing $\widetilde{F}=10^6$ Wm$^{-2}$.  (a) $\widetilde{F}_{\rm conv}$, (b) $\partial S/\partial r$, (c) $\widetilde{F}_{\rm mix}$, and (d) $S_{\rm liq}$.  Dashed line in (a) and (c) is the prescribed total flux $\widetilde{F}$.}
\label{fig:semianal}
\end{figure}

To elucidate this behaviour further it is useful to restrict the subsequent analysis to a range of $dS_{\rm liq}/dr$ that is physically reasonable.  The liquidus $S_{\rm liq}$ controls the pressure at which crystals first form in a magma ocean.  For the standard ``bottom-up'' crystallisation scenario, $dS_{\rm liq}/dr$ remains everywhere negative.  In contrast, a ``middle-out'' crystallisation scenario---recently proposed for the early Earth \citep{SKS09}---results from a liquidus overturn where the sign of the gradient changes at around 75 GPa (Fig.~\ref{fig:semianal}d).  Using this liquidus we solve the simplified model for a heat flux of 10$^6$ Wm$^{-2}$ (Fig.~\ref{fig:semianal}).  In addition to again revealing the large dynamic range of $\partial S/\partial r$ (Fig.~\ref{fig:semianal}b), we see that heat is dominantly transported by mixing where $dS_{\rm liq}/dr>0$ (Fig.~\ref{fig:semianal}c).  For $dS_{\rm liq}/dr<0$, however, both convection and mixing flux have large magnitude ($\sim$10$^{12}$ Wm$^{-2}$) and opposite signs that largely cancel to result in a smaller positive flux of $10^6$ Wm$^{-2}$ (Fig.~\ref{fig:semianal}a,c).

This analysis provides crucial insight for devising a suitable numerical method for the full system of equations.  Importantly, we must ensure the numerical method is both accurate and stable when $\partial S/\partial r$ varies over 12 orders of magnitude.  Furthermore, $\partial S/\partial r$ appears inside a non-linear term (square root) which can amplify errors when computing fluxes.  This is because the square root function has a vertical tangent line at the origin, so as $\partial S/\partial r \rightarrow 0$ a proportionally small change of $\partial S/\partial r$ results in a large relative change of flux.  The total error can be further compounded if component fluxes are computed individually and differenced to obtain the total flux; this is because for $dS_{\rm liq}/dr<0$ the total flux arises from the combination of a positive convective flux and a negative mixing flux, yielding $\widetilde{F} \sim 10^6$ Wm$^{-2}$, which is around 6 orders of magnitude less than both $\widetilde{F}_{\rm conv}$ and $\widetilde{F}_{\rm mix}$ (Fig.~\ref{fig:semianal}c,d).  However, in this part of the domain the entropy gradient is comparatively large which will help to mitigate large fluctuations in flux due to small changes of $\partial S/\partial r$.  For $dS_{\rm liq}/dr>0$, the total flux is dominantly carried by mixing and hence there is less concern that differencing the convective and mixing flux will introduce significant error or loss of precision.  However, it is important to retain precision in the estimate of $\partial S/\partial r$ since the entropy gradient is generally small throughout convecting systems (Fig.~\ref{fig:semianal}b).
\section{Numerical method}
\label{sect:numerical}
\subsection{Auxiliary variable approach}
We use the finite volume method (FVM) on a fixed grid to formulate a numerical solution to the full model because it is naturally energy conserving, intuitive, and flexible.  It is straightforward to compute fluxes of arbitrary algebraic complexity in a FVM which is important because the fluxes are non-linear and depend on material parameters.  Furthermore, a FVM can easily accommodate additional fluxes with minor modifications to ensure our method remains customisable.  There are also fully dynamic mantle convection codes that use the FVM \citep[e.g,][]{TACK08} and hence we can readily port components of our model to other codes.

The simplified model (Section~\ref{sect:simplified}) reveals a large dynamic range for $\partial S/\partial r$ of around 12 orders of magnitude (Figs.~\ref{fig:regime}, \ref{fig:semianal}).  Therefore, controlling numerical errors in $\partial S/\partial r$ is important for computing an accurate and robust numerical solution.  For this reason we solve for the time-dependence of an auxiliary variable $q=\partial S/\partial r$, rather than $S$ as would typically be done, since this eliminates the need to compute a finite difference estimate of $\partial S/\partial r$ using $S$.  This is advantageous because $S$ is defined pointwise on a numerical mesh and near-identical neighbouring values are expected in regions where $dS_{\rm liq}/dr>0$ (Fig.~\ref{fig:semianal}).  In these regions the finite precision representation of floating point numbers can result in catastrophic cancellation and amplify numerical noise.  However, we do not eliminate finite difference operations entirely from the numerical method as they are still implicitly used to obtain a numerical estimate of $\partial q/\partial t$ (Eq.~\ref{eqn:ddSdrdt1}) and explicitly used to compute an approximate Jacobian (Section~\ref{sec:timeintegration}).
Using the auxiliary variable $q=\partial S/\partial r$ and equivalence of mixed partial derivatives (assuming $S$ is sufficiently smooth), it follows that:
\begin{equation}
\label{eqn:dqdt}
\frac{\partial q}{\partial t} = \frac{\partial}{\partial r}\frac{\partial S}{\partial t}
\end{equation}
Assuming radial symmetry and substituting this expression into Eq.~\ref{eqn:spherical_cons} gives the continuous form:
\begin{equation}
\label{eqn:dqdt_diff}
\frac{\partial q}{\partial t} = - \frac{\partial}{\partial r} \left[ \frac{1}{\rho T r^2} \frac{\partial}{\partial r} (r^2 F(q,S)) \right]
\end{equation}
However, we actually formulate and solve the problem using the integral form (Eq.~\ref{eqn:entropy_integral}) to determine $\partial S/\partial t$ and subsequently calculate $q$ using Eq.~\ref{eqn:dqdt}.  Both the continuous and integral solution approaches require an additional equation to relate $q$ and $S$:
\begin{equation}
\label{eqn:S_tie}
S(r,t) = S_{\frac{1}{2}}(r_0,t) + \int_{r_0}^r q(r,t) dr
\end{equation}
Eq.~\ref{eqn:S_tie} requires that the entropy is known at radius $r_0$ along the entropy profile ($S_\frac{1}{2}$) in order to unambiguously recover $S(r,t)$ using $q$; this is facilitated by Eq.~\ref{eqn:entropy_integral}.  The index of $\frac{1}{2}$ in Eq.~\ref{eqn:S_tie} is arbitrary but has been chosen for consistency with the semi-discrete equations that are presented in Section~\ref{sect:discrete}. 
\subsection{Semi-discretisation}
\label{sect:discrete}
We solve the energy transport equations (Eqs.~\ref{eqn:entropy_integral}, \ref{eqn:dqdt}, \ref{eqn:S_tie}) with spherically-symmetric geometry using the FVM to compute $S(r,t)$.  It is natural to employ a staggered grid where fluxes are defined at basic nodes at cell boundaries and quantities that are integrated over a control volume (cell) are associated with staggered nodes, which are defined as equidistant between neighbouring basic nodes.  We discretise the spatial coordinate (radius) using a piecewise constant reconstruction and evaluate integrals using the midpoint rule.  The energy balance (Eq.~\ref{eqn:entropy_integral}) can then be expressed as a non-linear system of equations to solve for $\partial q/\partial t$ (Eq.~\ref{eqn:dqdt}) at the basic nodes:
\begin{subequations}{\label{eqn:ddSdrdt}}
\begin{align}
\frac{\partial q_i}{\partial t} &= \frac{1}{\Delta r_i} \left( \frac{\partial S_{i+\frac{1}{2}}}{\partial t} - \frac{\partial S_{i-\frac{1}{2}}}{\partial t} \right) \label{eqn:ddSdrdt1}\\
&= \frac{1}{\Delta r_i} \left( \frac{{4 \pi r_{i+1}}^2\mathcal{F}_{i+1}}{\mathcal{C}_{i+\frac{1}{2}}} - \frac{4 \pi {r_i}^2 \mathcal{F}_i}{\mathcal{C}_{i+\frac{1}{2}}} - \frac{4 \pi{r_i}^2 \mathcal{F}_i}{\mathcal{C}_{i-\frac{1}{2}}} +  \frac{4 \pi {r_{i-1}}^2\mathcal{F}_{i-1}}{\mathcal{C}_{i-\frac{1}{2}}} \right. \nonumber \\
& \qquad \qquad + \left. \frac{H_{i+\frac{1}{2}}}{T_{i+\frac{1}{2}}} - \frac{H_{i-\frac{1}{2}}}{T_{i-\frac{1}{2}}} \right) \label{eqn:ddSdrdt2}
\end{align}
\end{subequations}
and $\partial S/\partial t$ at the uppermost staggered node:
\begin{equation}
\label{eqn:dS0dt}
\frac{\partial S_\frac{1}{2}}{\partial t} = \frac{4 \pi ({r_1}^2 \mathcal{F}_1-{r_0}^2 \mathcal{F}_0)}{\mathcal{C}_\frac{1}{2}} + \frac{H_\frac{1}{2}}{T_\frac{1}{2}}
\end{equation}
where $i$ is a mesh index that is zero at the top surface (a basic node) and $p$ at the bottom surface, $\mathcal{C}_i= \rho_i T_i V_i$ is the capacitance term on the LHS of Eq.~\ref{eqn:entropy_integral}, and $\mathcal{F}$ is the numerical flux which is an approximation to the true flux $F$ (Eq.~\ref{eqn:heat_flux}).  The discrete radial increment $\Delta r_i = r_{i+\frac{1}{2}} - r_{i-\frac{1}{2}}$ is negative by convention because the mesh index increases as radius decreases.  Eq.~\ref{eqn:ddSdrdt2} is recognised as a central finite difference of fluxes evaluated at cell boundaries (basic nodes) that are weighted by capacitances determined at cell centres (staggered nodes).  Since we solve for $q$ at the basic nodes we avoid a finite difference approximation for $\partial S/\partial r$ using $S$ at neighbouring staggered nodes.  This limits our scheme to just one finite difference operation acting on fluxes which also helps to retain overall numerical precision.  By tracking the entropy at the uppermost staggered node (Eq.~\ref{eqn:dS0dt}) we can integrate $q$ to obtain entropy $S(r,t)$ and therefore evaluate material properties at the basic and staggered nodes for $\mathcal{F}$ and $\mathcal{C}$, respectively (Eq.~\ref{eqn:S_tie}).  This integration is a numerically stable operation, and furthermore $S$ is used to return material quantities from smoothly-varying lookup tables (Section~\ref{sect:material}).  Therefore errors in $S$ are not amplified like errors in $\partial S/\partial r$.

We implement the boundary conditions (Section~\ref{sect:bc}) in our semi-discrete system of equations as follows.  The radiative surface boundary condition (Eq.~\ref{eqn:bc_radiative}) is:
\begin{equation}
\mathcal{F}_0 = \epsilon \sigma ({T_0}^4 - {T_{\rm eq}}^4)
\end{equation}
and using Eq.~\ref{eqn:bc_utbl}:
\begin{equation}
\label{eqn:bc_discrete}
T_\frac{1}{2} = T_0 + b {T_0}^3 
\end{equation}
We can solve for $T_0$ by finding the root of the cubic equation (Eq.~\ref{eqn:bc_discrete}) since $T_\frac{1}{2}$ is known at the current time and $b$ is constant.  The CMB condition (Eq.~\ref{eqn:bc_cmb}) is cast into an alternative form.  Recall that it is natural to formulate this condition using temperature as opposed to entropy since the core and mantle are thermally coupled.  We determine the thermal energy balance of the cell that neighbours the core:
\begin{equation}
V_{p-\frac{1}{2}} \rho_{p-\frac{1}{2}} c_{p-\frac{1}{2}} \frac{dT_{p-\frac{1}{2}}}{dt} = -4 \pi ({r_{p-1}}^2 \mathcal{F}_{p-1} - {r_p}^2 \mathcal{F}_p  )
\end{equation}
The lowermost cell is at the CMB temperature ($T_{p-\frac{1}{2}}=T_{\rm cmb}$) so we substitute in Eq.~\ref{eqn:bc_cmb} and rearrange to determine the CMB heat flux:
\begin{equation}
\label{eqn:bc_cmb_discrete}
\mathcal{F}_p = \left( \frac{r_{p-1}}{r_p} \right)^2 \left( 1+ \frac{V_{p-\frac{1}{2}} \rho_{p-\frac{1}{2}} c_{p-\frac{1}{2}}}{m_{\rm core} c_{\rm core} \hat{T}_{\rm core}} \right)^{-1} \mathcal{F}_{p-1}
\end{equation}
$\mathcal{F}_{p-1}$ is evaluated using Eq.~\ref{eqn:heat_flux} since it is a basic node within the magma ocean solution domain.  Note that $\rho_{p-1/2}$ and $c_{p-1/2}$ depend on entropy and pressure and hence are time-dependent.  Cast in this form it is readily apparent that the boundary condition provides no restriction on the heat flux that the core provides the mantle.  It effectively enables the (unmodelled) ultra-thin thermal boundary layer at the base of the mantle to instantaneously adjust its thickness to accommodate the magma ocean heat flux.
\subsection{Time integration}
\label{sec:timeintegration}
We formulate the numerical problem using C and \PETSc{} \citep{PETScman} and solve the resulting system of equations using \SUNDIALS{} \citep{HBG05,HS15}. \SUNDIALS{}'s and \PETSc{}'s interfaces are customised to facilitate quadruple precision (\texttt{\_\_float128} from GCC libquadmath) calculations and to use CVODE's direct sequential linear solver.  We use the stiff ODE solver (CVODE) in the \SUNDIALS{} package, based on backward differentiation formulae.  This implicit timestepping approach is warranted because the system is stiff owing to the large difference in timescales between inviscid and viscous flow.  We configure the ODE solver to use a 5th order method with dynamic time stepping and automatic order reduction if the solution becomes unstable.

Since the problem has low dimensionality, we use a direct solve within the Newton iteration, rather than an iterative approach, thus avoiding introducing additional error in the form of a convergence tolerance.  We opt to compute a finite-difference Jacobian, due to the non-trivial nature of the fluxes that depend on gradients and also material properties obtained from lookup tables; in practice these are often obtained from opaque third-party software which does not provide derivatives.
\section{Results}
\label{sect:results}
\begin{figure}[t]
\centering
\includegraphics[width=0.8\textwidth]{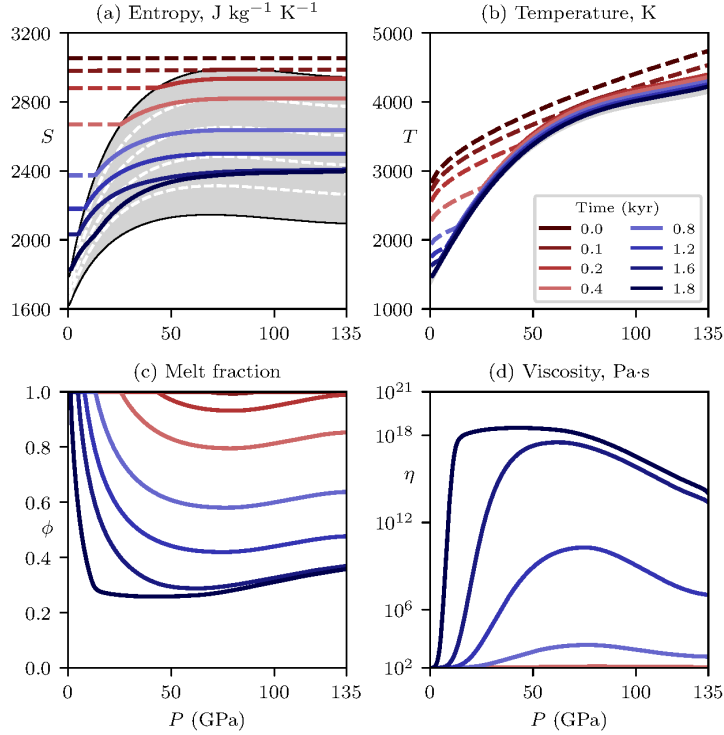}
\caption{Evolution (0--1.8 kyrs) of case MO where convective heat transport and mixing control the evolution prior to reaching the rheological transition.  The surface and CMB are at 0 and 135 GPa, respectively, and the mantle is initially superliquidus everywhere (0 kyr).  (a) Entropy.  The mixed phase region is grey, bounded by the liquidus above and solidus below, with melt fraction ($\phi$) contours every 0.2 units denoted by white dashed lines.  Lines are solid for the mixed phase and dashed for the pure liquid.  (b) Temperature, (c) Melt fraction, and (d) Viscosity.}
\label{fig:middle-out-1}
\end{figure}

We demonstrate our numerical method by investigating two crystallisation scenarios for Earth that are dictated by the shape of the liquidus and solidus.  The liquidus in case ``MO'' (middle-out) is derived from \cite{SKS09} and has a characteristic overturn in the middle of the domain and hence $dS_{\rm liq}/dr$ changes sign from negative to positive as pressure increases (Fig.~\ref{fig:middle-out-1}a).  In contrast, case ``BU'' (bottom-up) uses the liquidus from \cite{ABL11} where crucially $dS_{\rm liq}/dr<0$ everywhere (Fig.~\ref{fig:bottom-up-1}a).  These archetypal cases demonstrate the role of $dS_{\rm liq}/dr$ in dictating the necessary numerical precision to obtain a satisfactory solution.  All other physical parameters are identical (Table~\ref{tbl:parameters}) and we use a regular grid with 200 basic nodes ($p=200$).

\begin{table}[t]
\centering
\begin{tabular}{| l | l | l | l |} 
  \hline
Parameter & Symbol & Value & Units \\
\hline
Core-mantle boundary radius & $r_{\rm cmb}$ & $3.504 \times 10^6$ & m\\
Core heat capacity & $c_{\rm core}$ & 880 & J kg$^{-1}$ K$^{-1}$\\
Core mass & $m_{\rm core}$ & $1.935 \times 10^{24}$ & kg\\
Core cooling parameter & $\hat{T}_{\rm core}$ & 1.147 &\\
Critical Reynolds number & $Re_{\rm crit}$ & 9/8 &\\
Eddy diffusivity & $\kappa_h$ & Eq.~\ref{eqn:kappah} & m$^2$ s$^{-1}$\\
Emissivity & $\epsilon$ & 1.0 &\\
Entropy & $S$ & Eq.~\ref{eqn:entropy_integral} & J kg$^{-1}$ K$^{-1}$\\
Equilibrium temperature & $T_{\rm eq}$ & 273 & K\\
Grain size & $a$ & $10^{-3}$  & m\\
Gravity & $g$ & -10 & m s$^{-2}$\\
Internal heat generation & $H$ & 0 & W kg$^{-1}$\\
Inviscid flow velocity & $u_{\rm invis}$ & Eq.~\ref{eqn:uinvis} & m s$^{-1}$\\
Thermal boundary layer scaling & $b$ & $10^{-7}$ & K$^{-2}$\\
Thermal conductivity & $k$ & 4.0 & W m$^{-1}$ K$^{-1}$\\
Viscous flow velocity & $u_{\rm visc}$ & Eq.~\ref{eqn:uvisc} & m s$^{-1}$\\
Hydrostatic pressure & $P$ & Eq.~\ref{eqn:hydro_pressure} & Pa\\
\quad Density & $\rho_r$ & 4079 & kg m$^{-3}$\\
\quad Compressibility parameter & $\beta$ & $1.112 \times 10^{-7}$ & m$^{-1}$\\
Material properties (from lookup) &&&\\
\quad Adiabatic temperature gradient & $(dT/dP)_S$ & $f(S,P)$ & K Pa$^{-1}$\\
\quad Density & $\rho$ & $f(S,P)$ & kg m$^{-3}$\\
\quad Heat capacity & $c$ & $f(S,P)$ & J kg$^{-1}$ K$^{-1}$\\
\quad Temperature & $T$ & $f(S,P)$ & K\\
\quad Thermal expansion & $\alpha$ & $f(S,P)$ & K$^{-1}$\\
Viscosity & $\eta$ & Eq.~\ref{eqn:visc_nonskew} & Pa s\\
\quad Melt & $\eta_m$ & $10^2$ & Pa s\\
\quad Solid & $\eta_s$ & $10^{21}$ & Pa s\\
\quad Critical melt fraction & $\phi_c$ & 0.4 &\\
\quad Transition width & $\phi_w$ & 0.15 &\\
\hline
\end{tabular}

\caption{Parameters for the full model in cases MO and BU.}
\label{tbl:parameters}
\end{table}

\begin{figure}[th]
\centering
\includegraphics[width=0.8\textwidth]{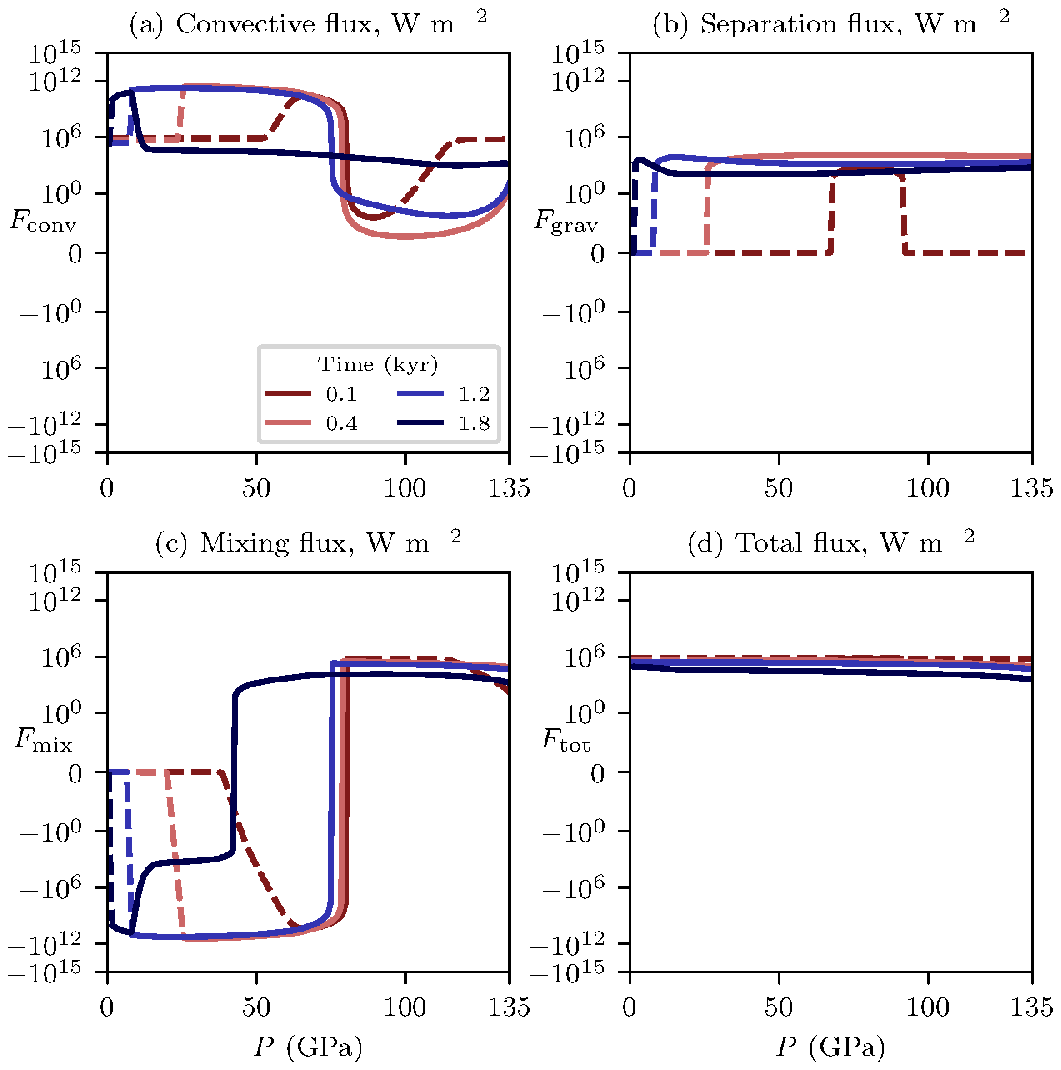}
\caption{Heat fluxes for case MO (Fig.~\ref{fig:middle-out-1}).  (a) Convection, (b) Gravitational separation, (c) Mixing, and (d) Total.  Note arcsinh transform for the y-axes.}
\label{fig:middle-out-2}
\end{figure}

\begin{figure}[th]
\centering
\includegraphics[width=0.8\textwidth]{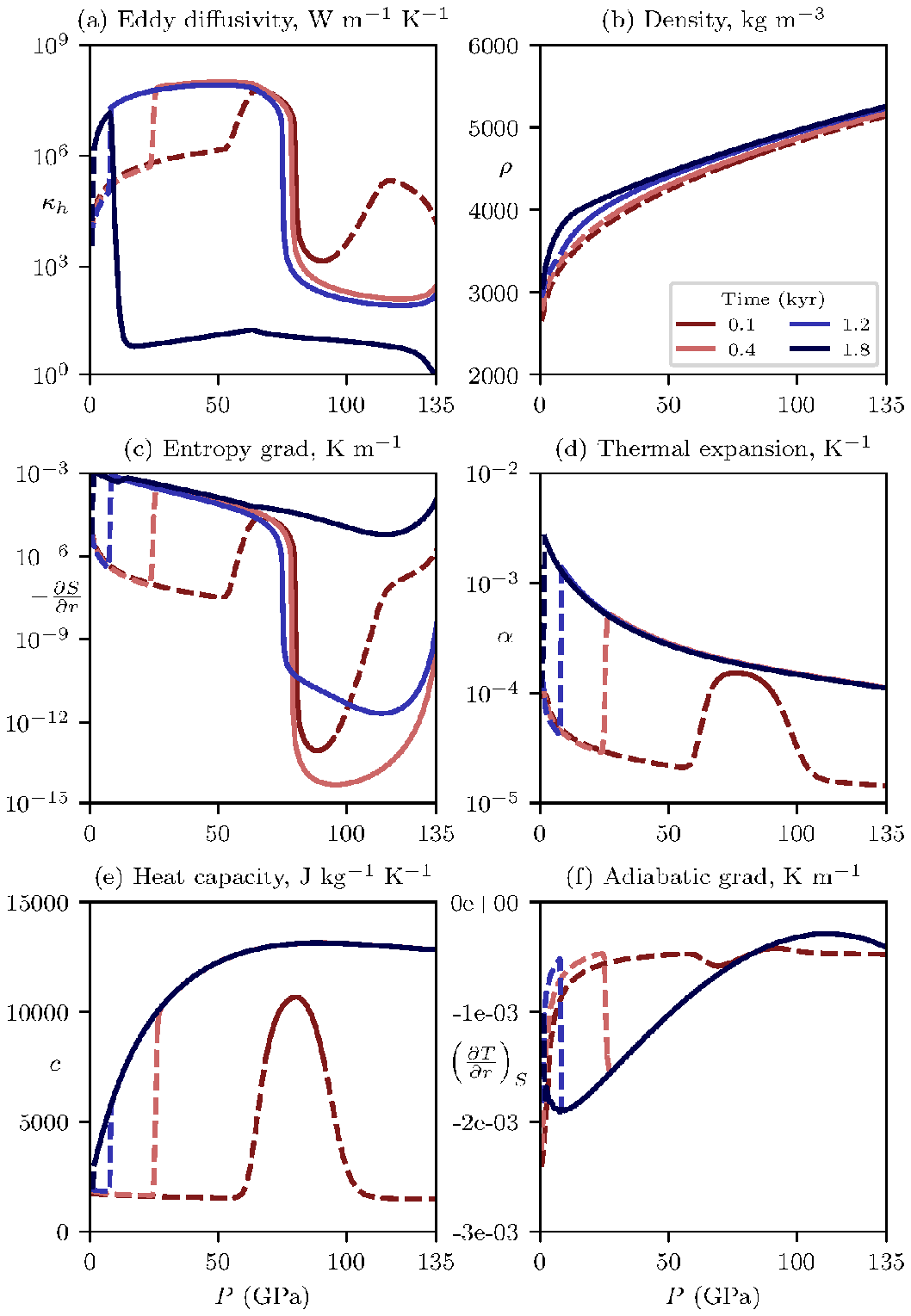}
\caption{Material properties for case MO (Fig.~\ref{fig:middle-out-1}).  (a) Eddy diffusivity, (b) Density, (c) Entropy gradient, (d) Thermal expansion coefficient, (e) Heat capacity, and (f) Adiabatic temperature gradient.}
\label{fig:middle-out-3}
\end{figure}

\begin{figure}[t]
\centering
\includegraphics[width=0.8\textwidth]{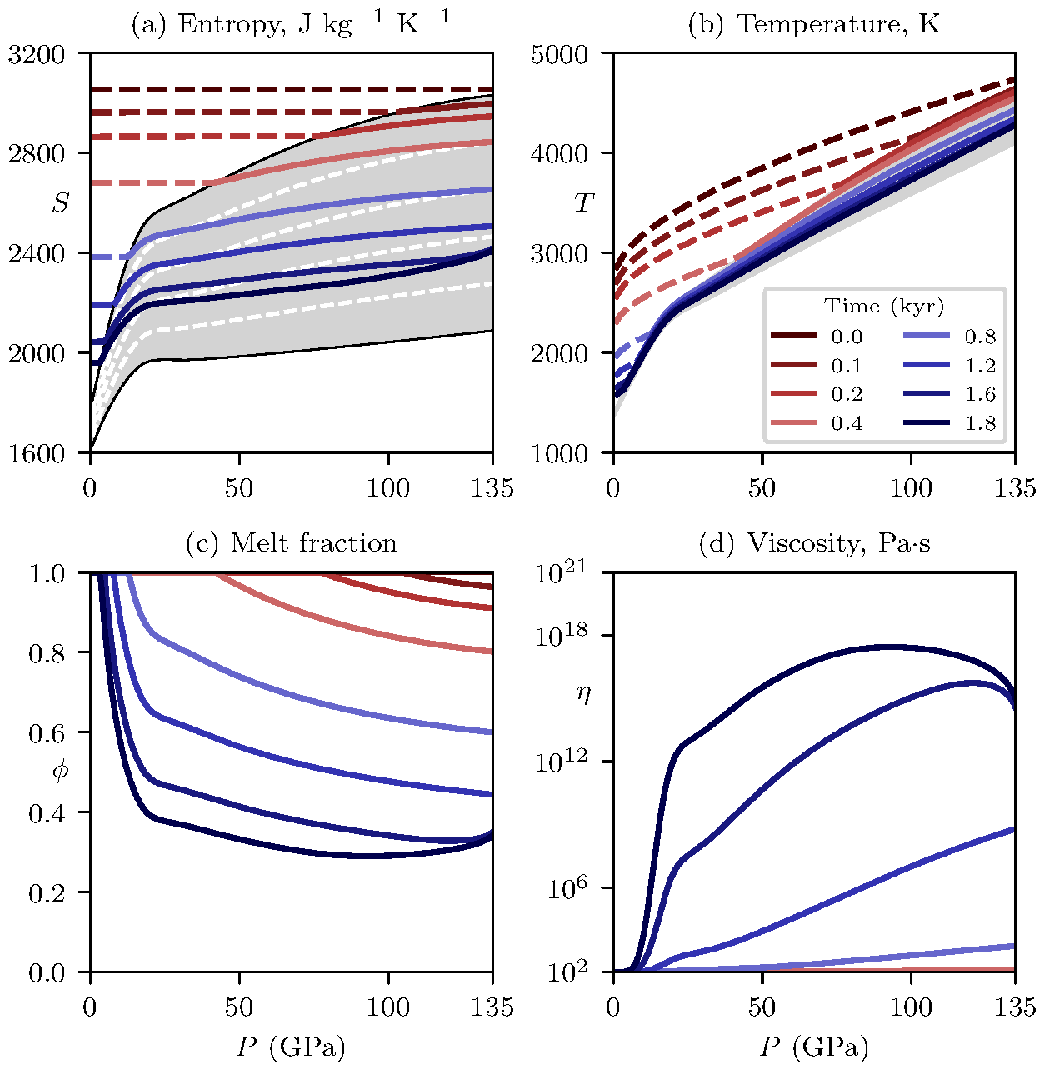}
\caption{Evolution (0--1.8 kyrs) of case BU.  (a) Entropy, (b) Temperature, (c) Melt fraction, and (d) Viscosity.  See Fig.~\ref{fig:middle-out-1} caption.}
\label{fig:bottom-up-1}
\end{figure}

For case MO we use quadruple precision calculations with a relative and absolute timestepper tolerance of $10^{-18}$.  The mantle cools along approximate adiabats (constant $S$) until the adiabat intercepts the liquidus around 75 GPa (Fig.~\ref{fig:middle-out-1}a).  Then $\partial S/\partial r$ proceeds to decrease by 12 orders of magnitude below the liquidus as a consequence of $dS_{\rm liq}/dr$ switching sign from negative to positive.  Note the excellent agreement between the simplified and full models near the liquidus (compare Fig.~\ref{fig:semianal}b and Fig.~\ref{fig:middle-out-3}c, 0.4 kyr).  The large dynamic range of $\partial S/\partial r$ is the fundamental reason why quadruple precision is necessary for this case.  Where $dS_{\rm liq}/dr>0$, $\partial S/\partial r$ is driven to a small value ($\approx -10^{-13}$) to reduce the eddy diffusivity (Fig.~\ref{fig:middle-out-3}a) and hence the convective flux (Fig.~\ref{fig:middle-out-2}a).  This is because the majority of the total flux is accommodated by mixing (Eq.~\ref{eqn:Fmix}, Fig.~\ref{fig:middle-out-2}c) since $\partial \phi/\partial r<0$ in this region ($\kappa_h>0$).  This behaviour adheres to the insight gained from the simplified model (Section~\ref{sect:simplified}) where small negative $\partial S/\partial r$ is expected for $dS_{\rm liq}/dr>0$.  Despite the large variation in convective and mixing fluxes across the mantle, the total flux is near-constant and decreases steadily with time (Fig.~\ref{fig:middle-out-2}d).

\begin{figure}[t]
\centering
\includegraphics[width=0.8\textwidth]{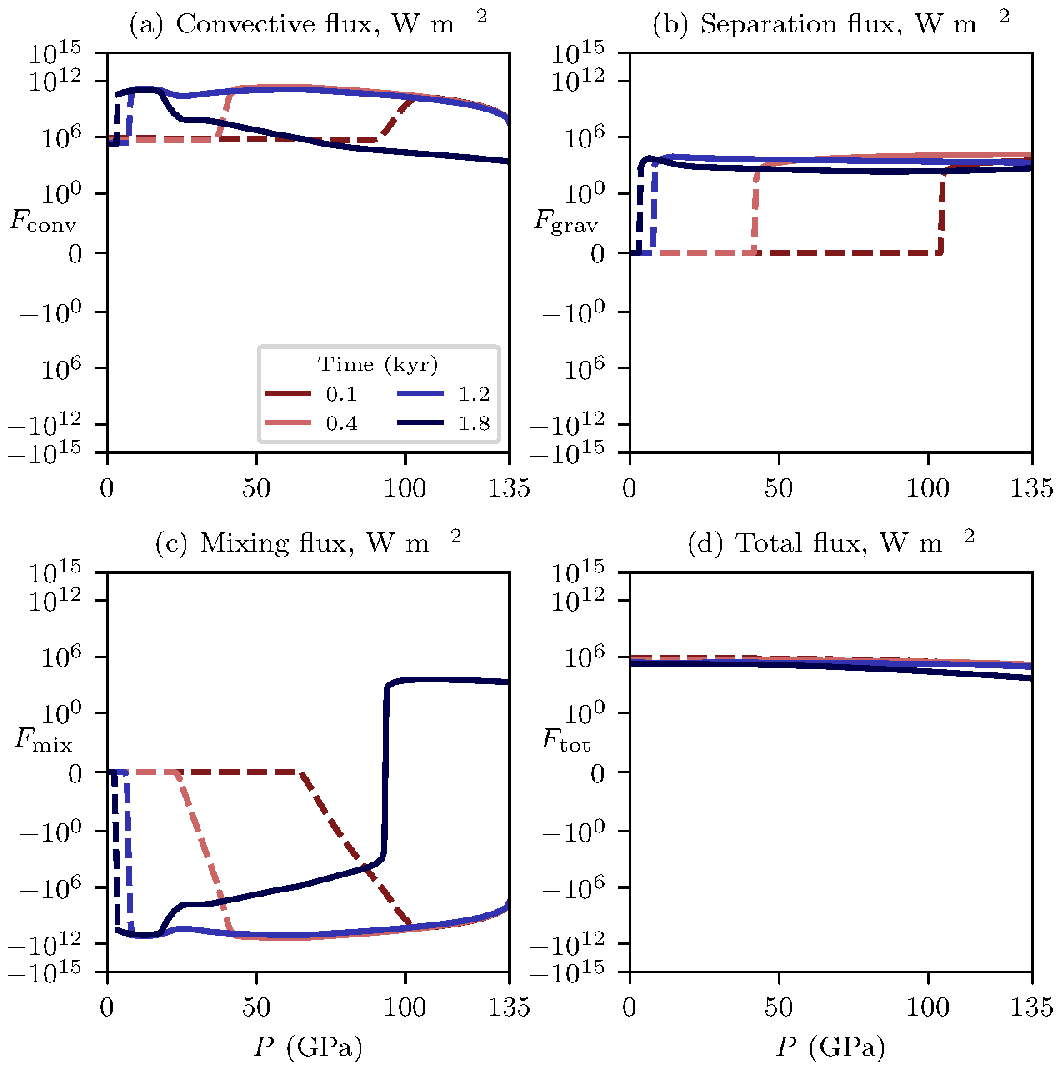}
\caption{Heat fluxes for case BU (Fig.~\ref{fig:bottom-up-1}).  (a) Convection, (b) Gravitational separation, (c) Mixing, and (d) Total.  Note arcsinh transform for the y-axes.}
\label{fig:bottom-up-2}
\end{figure}

\begin{figure}[t]
\centering
\includegraphics[width=0.8\textwidth]{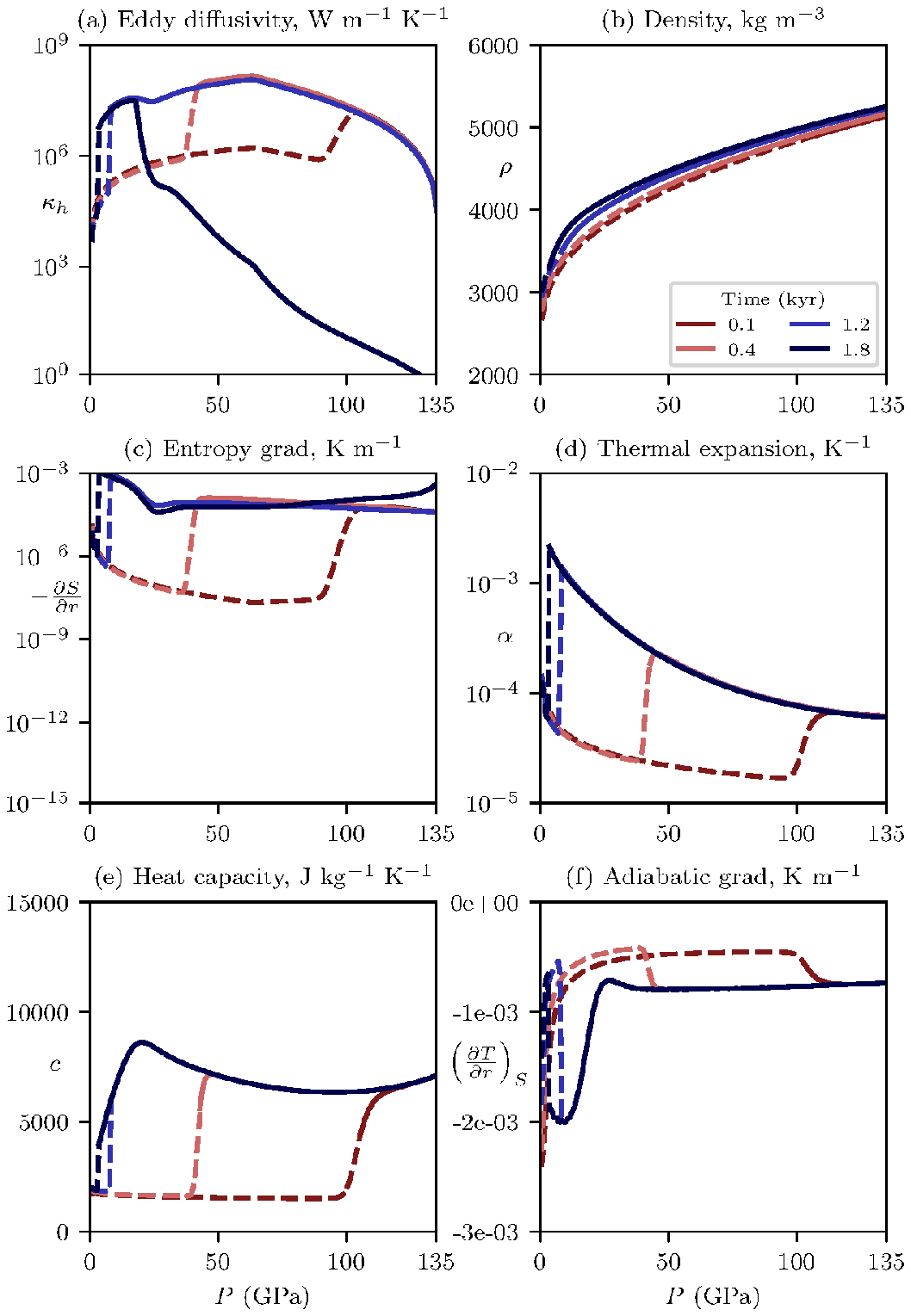}
\caption{Material properties for case BU (Fig.~\ref{fig:bottom-up-1}).  (a) Eddy diffusivity, (b) Density, (c) Entropy gradient, (d) Thermal expansion coefficient, (e) Heat capacity, and (f) Adiabatic temperature gradient.}
\label{fig:bottom-up-3}
\end{figure}

For case BU, $\partial S/\partial r$ remains relatively large (negative) and crucially has a reduced dynamic range in comparison to MO (Fig.~\ref{fig:bottom-up-3}c).  As the mantle cools, the liquidus is first intercepted by the adiabat at the bottom of the domain (Fig.~\ref{fig:bottom-up-1}a).  The simplified model reveals that $\partial S/\partial r$ is always less than half of $dS_{\rm liq}/dr$ (Eq.~\ref{eqn:Fg0}), which restricts the dynamic range of $\partial S/\partial r$.  This ensures that the system is resolvable using double precision calculations and we set the relative and absolute tolerance of the time-stepper to $10^{-10}$.  Prior to 1.2 kyr, convection (Fig.~\ref{fig:bottom-up-2}a) and mixing (Fig.~\ref{fig:bottom-up-2}c) have comparable magnitude but opposite sign, which drives a net flux of around 10$^6$ Wm$^{-2}$ (Fig.~\ref{fig:bottom-up-2}d).  At later time (1.8 kyr), however, the mixing flux changes sign in the deepest part of the mantle (Fig.~\ref{fig:bottom-up-2}c).  This is because $\partial \phi/\partial r$ switches sign as a consequence of the cooling profile lying over the rheological transition.  Note that this sign change does not occur for MO because $\partial \phi/\partial r<0$ is already established in the lower mantle (Fig.~\ref{fig:middle-out-1}c).  The total variation of the mixing flux for BU is not as large as for MO (compare Fig.~\ref{fig:middle-out-2}c and Fig.~\ref{fig:bottom-up-2}c).  But most importantly, $\partial S/\partial r$ remains relatively large for BU yet varies over about 12 orders of magnitude for MO (compare Fig.~\ref{fig:middle-out-3}c and Fig.~\ref{fig:bottom-up-3}c).  This again emphasises the fundamental difference between the cases and the requirement for extended numerical precision for case MO.

\section{Discussion}
Our equations are based on \cite{ABE93,ABE95,ABE97} (Hereinafter ``Abe'') which uses mixing length theory (MLT) to formulate the convective flux in terms of an eddy diffusivity that is a function of the local super-adiabatic temperature gradient (Eq.~\ref{eqn:kappah}).  This follows a classic approach in stellar structure modelling where the mixing length explicitly appears in the equations and scales the gradient of entropy or temperature \citep[e.g.,][]{S71}, introducing the dependence of eddy diffusivity on $\partial S/\partial r$.  Furthermore, Abe's are the only studies (prior to this work) that consider convection, conduction, mixing, and gravitational separation---other studies typically just model the convective flux.  Abe uses a finite difference scheme with a modified Euler backward method of time stepping.  However, because he investigates bottom-up crystallisation ($dS_{\rm liq}/dr<0$ everywhere) he is not exposed to the numerical precision challenges that we uncover when the liquidus exhibits an overturn (case MO).  In fact, all previous work has focussed on bottom-up crystallisation similar to case BU where $dS_{\rm liq}/dr\ll0$ everywhere in the domain \citep[e.g.,][]{ABE93,ABE97,LMC13,MAS16}.

An alternative approach to formulate the convective flux, as opposed to MLT, uses boundary layer theory (BLT).  Formally, BLT is a simplified variant of MLT \citep{S94,K62} but for clarity we distinguish it separately to avoid confusion with our formulation based on Abe.  Analysis of boundary layer stability in Rayleigh-B\'{e}nard convection provides the classical result that the non-dimensional heat flux (Nusselt number) scales with the Rayleigh number to the power of one-third \citep[e.g.,][]{S94}.  In this case the length scale of convection is encapsulated within the Rayleigh number rather than a mixing length parameter.  The one-third scaling law is in reasonable agreement with experiments \citep[e.g., Table 1 in][]{GL00} and MLT \citep[see Fig. 3 in][]{ABE95}.

\cite{SS93} popularised BLT for application to magma oceans.  Subsequent studies that focus on understanding the energetic coupling and volatile exchange between a magma ocean and an atmosphere use BLT to quantify the convective heat transport in the interior \citep{LMC13,HAG13,HKA15}.  But since these studies focus on the role of the atmosphere, they implement a simple model of interior dynamics.  For example, they solve a single equation that describes the evolution of the mantle potential temperature, which involves calculating an effective heat capacity for the entire planet by using the mantle potential temperature to reconstruct the temperature profile as a function of depth.  This requires additional assumptions, most commonly that the mantle is adiabatic and predefining a path of crystallisation (e.g., bottom-up in the case of \cite{LMC13}).

Whilst the aforementioned quasi 1-D interior models inform about the bulk cooling of a planet, they provide limited information about the depth-dependence of the evolving interior.  This is where a local description of energy transfer within an interior is advantageous \citep[this study,][]{ABE93,MAS16}, since the evolving energy balance can be determined self-consistently with thermodynamic models of mantle materials \citep[e.g.,][]{WB17}.  For example, both case MO and BU demonstrate that the thermal structure in a crystallising magma ocean is not strictly adiabatic due to the latent heat transport associated with convective mixing.  A local description also provides flexibility to include additional energy fluxes and accommodate arbitrary liquidus and solidus curves.  This provides crucial infrastructure to investigate magma ocean crystallisation for rocky planets in general where cooling could be markedly different from Earth depending on planet composition, core size, atmospheric composition and dynamics, etc.

The convective flux determined by BLT can be used to compute an eddy diffusivity and may be formulated piecewise to account for regime changes such as a transition from soft to hard turbulence \citep{NBS14,MAS16}.  Since in BLT the convective flux is a power law of the Rayleigh number, the latter of which contains a difference of two temperatures, the functional complexity of the eddy diffusivity is considerably reduced for BLT in comparison to MLT.  This is because both the flux and hence eddy diffusivity do not depend on local gradients, which therefore reduces the requirement for precise numerical estimates of derivative quantities.  \cite{MAS16} determine a local Rayleigh number based on the difference between the temperature profile of the magma ocean and a reference adiabat, the latter of which is computed with a potential temperature corresponding to the surface temperature.  Hence although the flux is determined locally at each depth in the magma ocean, it is computed relative to a somewhat arbitrary reference adiabat---it is not apparent how physically reasonable this approach is.  They also ignore convective mixing and consider only bottom-up crystallisation, which alleviates any potential difficulties relating to the numerical precision of their finite difference scheme.

We choose the mixing length to be the distance to the nearest material boundary \citep[e.g.][]{SC97}, although alternative formulations exist (see supplementary material).  This formally means that during magma ocean crystallisation, whether from the bottom-up or middle-out, thermal boundary layers exist at the top and bottom of the mantle and hence single layer convection ensues.  The model thus presumes that crystal growth is sufficiently disrupted by turbulent flow that the magma ocean does not partition into multiple convecting domains.  Future work should investigate how to relax this requirement despite several physical and numerical challenges.  Imposing an abrupt change in the mixing length during crystallisation to allow formation of an additional boundary layer will likely produce artefacts that hinder physical interpretation of the model.  Furthermore, boundary layers in the magma ocean may only be a few centimetres thick and will migrate as crystallisation proceeds --- models with both high mesh resolution and a moving mesh may be required to address this scenario.

The angular momentum of the Earth-Moon system constrains the rotation rate of the early Earth to a few hours following the formation of the Moon.  Furthermore, accretion simulations predict that Earth-mass planets may initially rotate even faster, within about 30\% or less of their rotational stability limit \citep[e.g.,][]{KG10}.  This raises the question of how rotational effects can influence the subsequent cooling and crystallisation of a magma ocean.  In our modelling framework, the interaction of rotation and convection can be accommodated by modifying the mixing length.  \cite{KKS05} show that increasing the rotation rate reduces the spatial scale of convection and the efficiency of convective energy transport.  The mixing length decreases by a factor of 2 to 3 as the Rossby number decreases from 10 to 0.1.  These authors also find that the superadiabaticity increases as a function of latitude.
\section{Conclusions}
We present an intuitive and flexible method for solving a non-linear conservation law relevant to the interior dynamics of partially molten planets.  The numerical scheme solves the mixing length theory (MLT) representation of the energy balance.  In a single formulation (i.e., one solution domain with a fixed grid), our model captures both inviscid flow (high melt fraction) and viscous flow (low melt fraction) and thus can chart the cooling trajectory of an initially molten planet as it solidifies.  This encapsulates a viscosity range of 19 orders of magnitude.  Using the finite volume method we formulate a local description of the dynamics and consider heat transport due to conduction, convection, mixing (i.e., latent heat transport), and gravitational separation.  Our models are not constrained by \emph{a priori} assumptions that the magma ocean is necessarily convecting or that it will follow a particular crystallisation sequence (e.g., bottom-up).  It is straightforward to introduce additional energy transfer mechanisms within the finite volume framework.  The model interfaces with equations of state for melt and solid phases to ensure thermodynamic self-consistency for each phase and can additionally accommodate arbitrary melting curves.

Through a simplified MLT model, we reveal the requirement for extended numerical precision where the liquidus exhibits an overturn---a scenario which may be applicable to Earth.  In this case, it is necessary to resolve entropy gradients spanning 12 orders of magnitude.  We subsequently demonstrate the ability of the numerical scheme to handle this scenario using a full model which includes all four of the aforementioned energy fluxes, in addition to material properties obtained from lookup tables.  In contrast, the dynamic range of the entropy gradient is significantly less for a case where the liquidus monotonically increases with pressure, which explains why all previous modelling studies did not encounter the same technical challenge that we uncover.  Both the simplified and full model reveal cooling profiles that depart from adiabatic as a consequence of latent heat transport associated with convective mixing.  This emphasises the importance of considering latent heat transport in a magma ocean since this energy flux is frequently omitted in dynamic studies.

Finally, we expect the analysis and numerical scheme we propose to be of broad interest to other fields of science and engineering due to the prevalence of systems that are modelled by non-linear conservation laws.  In particular, we offer a simple and robust approach to accommodate a diffusion coefficient that is sensitive to a local gradient.
\section*{Code availability}
SPIDER: ``Simulating Planetary Interior Dynamics with Extreme Rheology'' is hosted on Bitbucket at \url{https://bitbucket.org/djbower/spider} and access can be requested by contacting Dan J. Bower.

\section*{Acknowledgements}
The ETH Z\"{u}rich Postdoctoral Fellowship Program is gratefully acknowledged for providing DJB with the opportunity and resources to conduct this research.  PS acknowledges financial support from the Swiss University Conference and the Swiss Council of Federal Institutes of Technology through the Platform for Advanced Scientific Computing (PASC) program.  ASW acknowledges the financial support and scientific freedom provided by the Turner Postdoctoral Fellowship at the University of Michigan, Ann Arbor.  We thank Paul J. Tackley for discussions and for providing comments on an earlier version of the manuscript.  Two anonymous reviewers provided helpful critique that enhanced the clarity of the manuscript.  Perceptually-uniform and colour-blindness friendly colour arrays are available at \url{http://www.fabiocrameri.ch/visualisation.php}.

\section*{References}
\bibliography{refs}

\clearpage

\FloatBarrier
\appendix

\section{Supplementary Material}

\section*{Demonstration of convergence}
Our numerical solution scheme uses the finite volume method (FVM) with an auxiliary variable to enable greater numerical precision to be achieved.  In this regard, there is extensive discussion in the literature on the accuracy and convergence behaviour of the FVM \citep[e.g.,][]{DJ14,LEV02};
we note that our introduction of an auxiliary variable can be seen simply as change of variables for the semi-discretised system.  Nevertheless, given the unusual nature of our system, particularly the dependence of a diffusion coefficient on a gradient, we provide numerical demonstrations of the convergence behaviour of our scheme here.  An estimate of the tightest upper bound of the total error incurred in a central finite difference is on the order of $10^{-11}$ using double precision calculations \citep[e.g.,][]{NW06}.  Hence this is the error we can expect when we evaluate $\partial q_i/\partial t$ using Eq.~32 for models that use double precision such as case BU.  Therefore, this constrains the tightest tolerance for the accuracy of the timestepper to also be around $10^{-11}$, on the basis that this is the maximum total accuracy we can expect to achieve.

\begin{figure}[t]
\centering
\begin{subfigure}{.5\textwidth}
  \centering
  \includegraphics[width=1.0\linewidth]{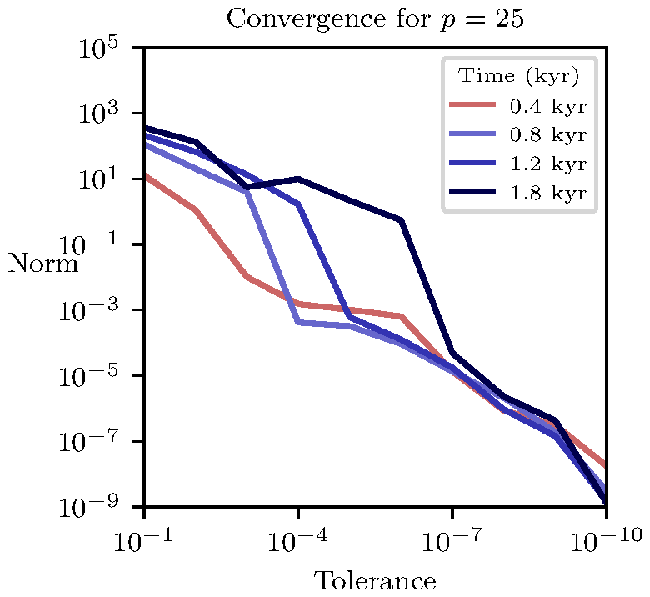}
  \caption{Convergence of BU for $p=25$}
  \label{fig:sub1}
\end{subfigure}%
\begin{subfigure}{.5\textwidth}
  \centering
  \includegraphics[width=1.0\linewidth]{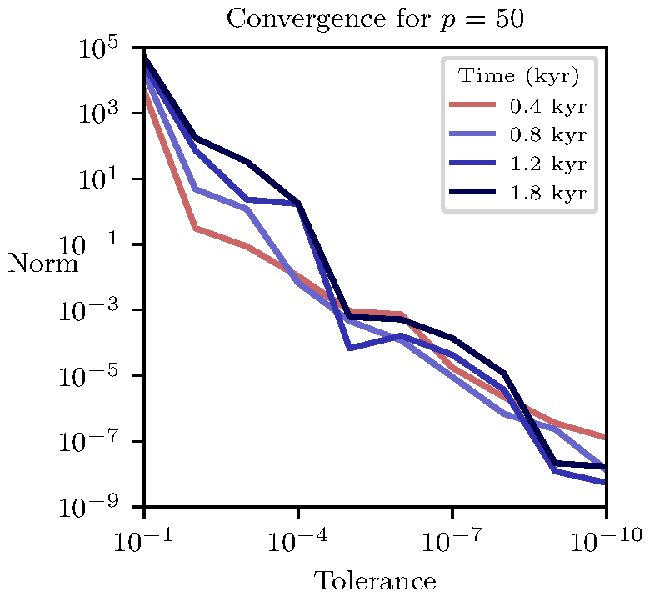}
  \caption{Convergence of BU for $p=50$}
  \label{fig:sub2}
\end{subfigure}\\~\\
\vspace{0.5cm}
\begin{subfigure}{.5\textwidth}
  \centering
  \includegraphics[width=1.0\linewidth]{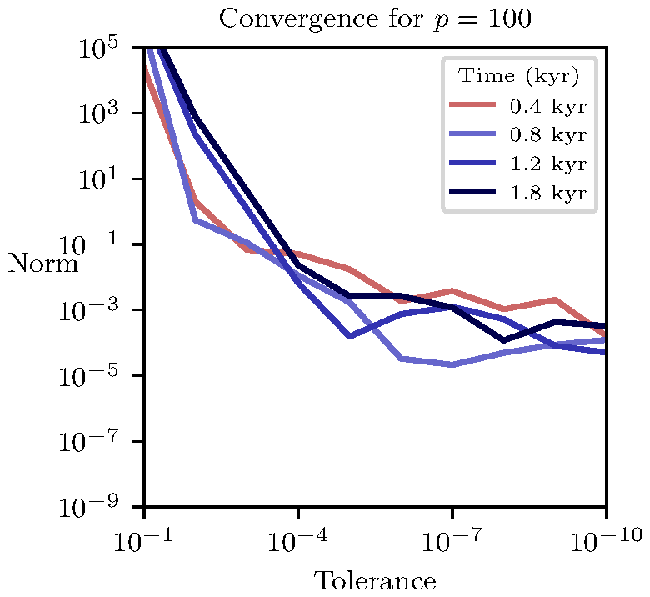}
  \caption{Convergence of BU for $p=100$}
  \label{fig:sub3}
\end{subfigure}%
\begin{subfigure}{.5\textwidth}
  \centering
  \includegraphics[width=1.0\linewidth]{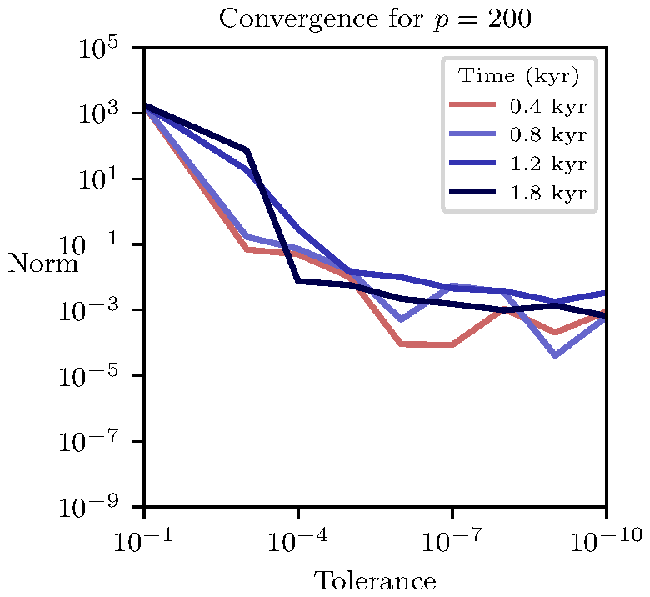}
  \caption{Convergence of BU for $p=200$}
  \label{fig:sub4}
\end{subfigure}
\caption{Convergence tests for case BU for (a) $p=25$, (b) $p=50$, (c) $p=100$, and (d) $p=200$, for the solution vectors corresponding to approximately 25\% (0.4 kyr), 50\% (0.8 kyr), 75\% (1.2 kyr) and 100\% (1.8 kyr) of the total integration time.  By definition, the norm of the smallest tolerance ($10^{-11}$) is zero.}
\label{fig:tol}
\end{figure}

To demonstrate convergence, we thus rerun case BU with the absolute and relative tolerances of the timestepper ranging from $10^{-1}$ to $10^{-11}$ for mesh sizes (number of basic nodes) of $p=$25, 50, 100, and 200.  The nominal case BU in the main manuscript has a tolerance of $10^{-10}$ and $p=200$.  For each $p$ we compute the Euclidean norm of the difference between the solution vector for a given tolerance and the vector for the tightest tolerance ($10^{-11}$) at four output times that roughly correspond to 25\%, 50\%, 75\% and 100\% of the total integration time.  The results are summarised in Fig.~\ref{fig:tol}.  Importantly, the norm generally decreases as the timestepper tolerance decreases for all mesh sizes considered, which is consistent with a convergent scheme.
The scheme contains several complicating factors, in addition to its inherent nonlinearities, including nonlinear boundary conditions and smoothing parameters (as discussed in the following section); thus, even though exact benchmark solutions or rigorous convergence analysis are not available, we are confident in the convergence of our scheme to a physically meaningful solution.

\section*{Sensitivity analysis}
\subsection*{Resolution of lookup tables}
Melt and solid thermophysical properties are stored in lookup tables that are accessed as the model timesteps.  For case BU in the main manuscript the lookup tables have a resolution of approximately 23 Jkg$^{-1}$K$^{-1}$ in entropy and 2 GPa in pressure.  We test the sensitivity of our result to this resolution by running an additional three cases that have the same parameters as BU and coarser lookup table resolutions: (1) Pressure data coarsened by a factor of two, (2) Entropy data coarsened by a factor of two, (3) Both pressure and entropy data coarsened by a factor of two.  In all three of these cases the difference relative to BU is visually imperceptible (Fig.~6) which suggests that the lookup tables have sufficient resolution to not influence our results.
\subsection*{Smoothing width $\phi_w$}
The lookup tables contain smooth data and the bilinear interpolation scheme in the code effectively provides additional smoothing.  Therefore, the only smoothing parameter that we implement is to ensure that material properties vary smoothly across the liquidus and solidus; a smoothing width $\phi_w$ and Eq.~15 are utilised for this cause.  Smoothing is required to ensure a stable numerical solution and it is important to note that the smoothing width does not correspond to a physical parameter.  In cases BU and MO in the main manuscript the smoothing width $\phi_w=0.01$ is 1\% of the width of the mixed phase region in (non-dimensional) units of melt fraction; clearly, $\phi_w$ should always be much smaller than the width of the mixed phase region.  To test the sensitivity of our result to this choice we rerun BU using a smoothing width 5 times larger ($\phi_w=0.05$) and 5 times smaller ($\phi_w=0.002$) than the nominal case.  The results of these two cases are near-visually identical to BU (Fig.~6) and are therefore omitted from this supplementary material.  Nevertheless, this analysis shows that the model results are robust for smoothing widths ranging by more than an order of magnitude.
\subsection*{Mixing length}
\begin{figure}[t]
\centering{
\includegraphics[width=0.8\textwidth]{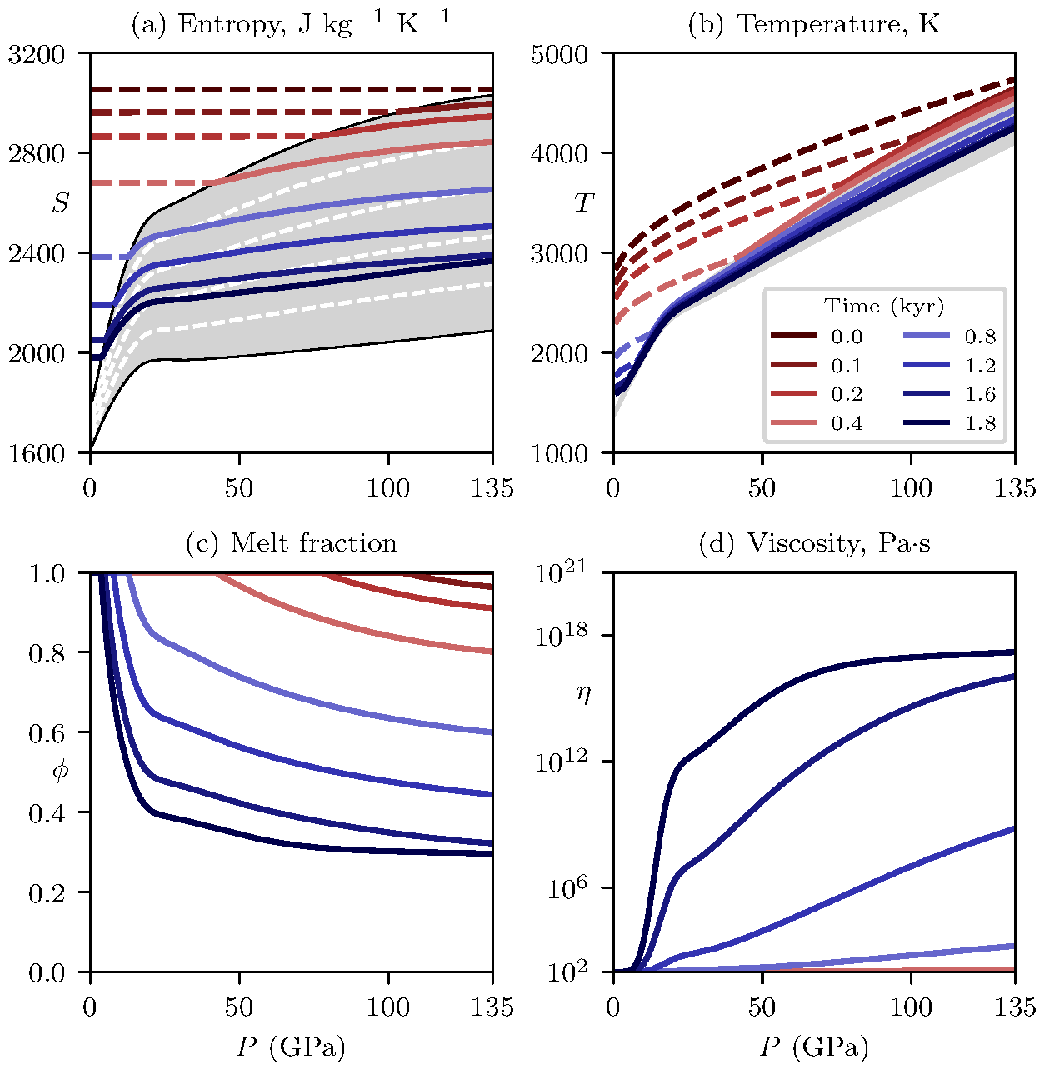}
}
\caption{Evolution (0--1.8 kyrs) of case BUM to compare with case BU (Fig.~6).  (a) Entropy, (b) Temperature, (c) Melt fraction, and (d) Viscosity.}
\label{fig:bottom-up-modified-1}
\end{figure}

\begin{figure}[t]
\centering{
\includegraphics[width=0.8\textwidth]{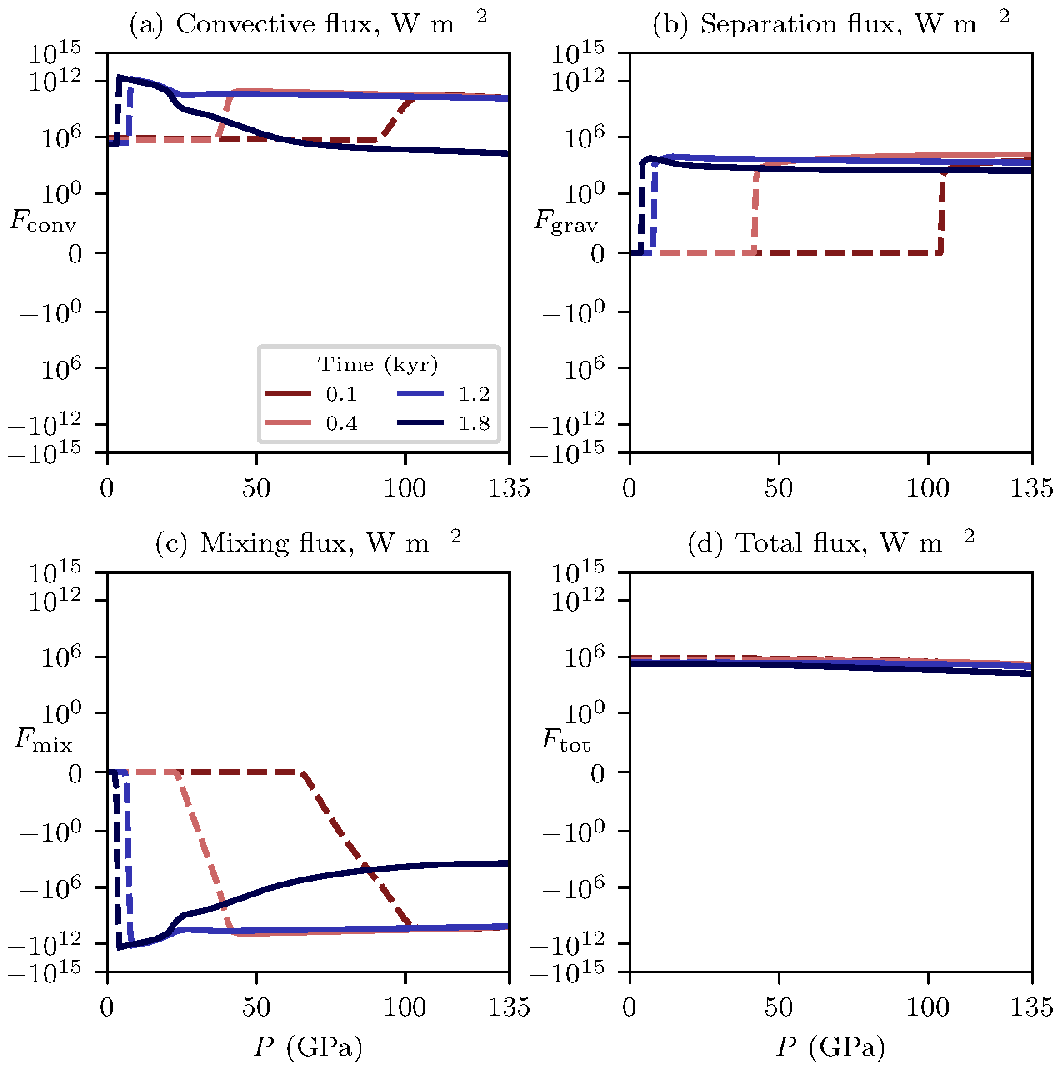}
}
\caption{Heat fluxes for case BUM (Fig.~\ref{fig:bottom-up-modified-1}) to compare with case BU (Fig.~7).  (a) Convection, (b) Gravitational separation, (c) Mixing, and (d) Total.  Note arcsinh transform for the y-axes.}
\label{fig:bottom-up-modified-2}
\end{figure}

\begin{figure}[t]
\centering{
\includegraphics[width=0.8\textwidth]{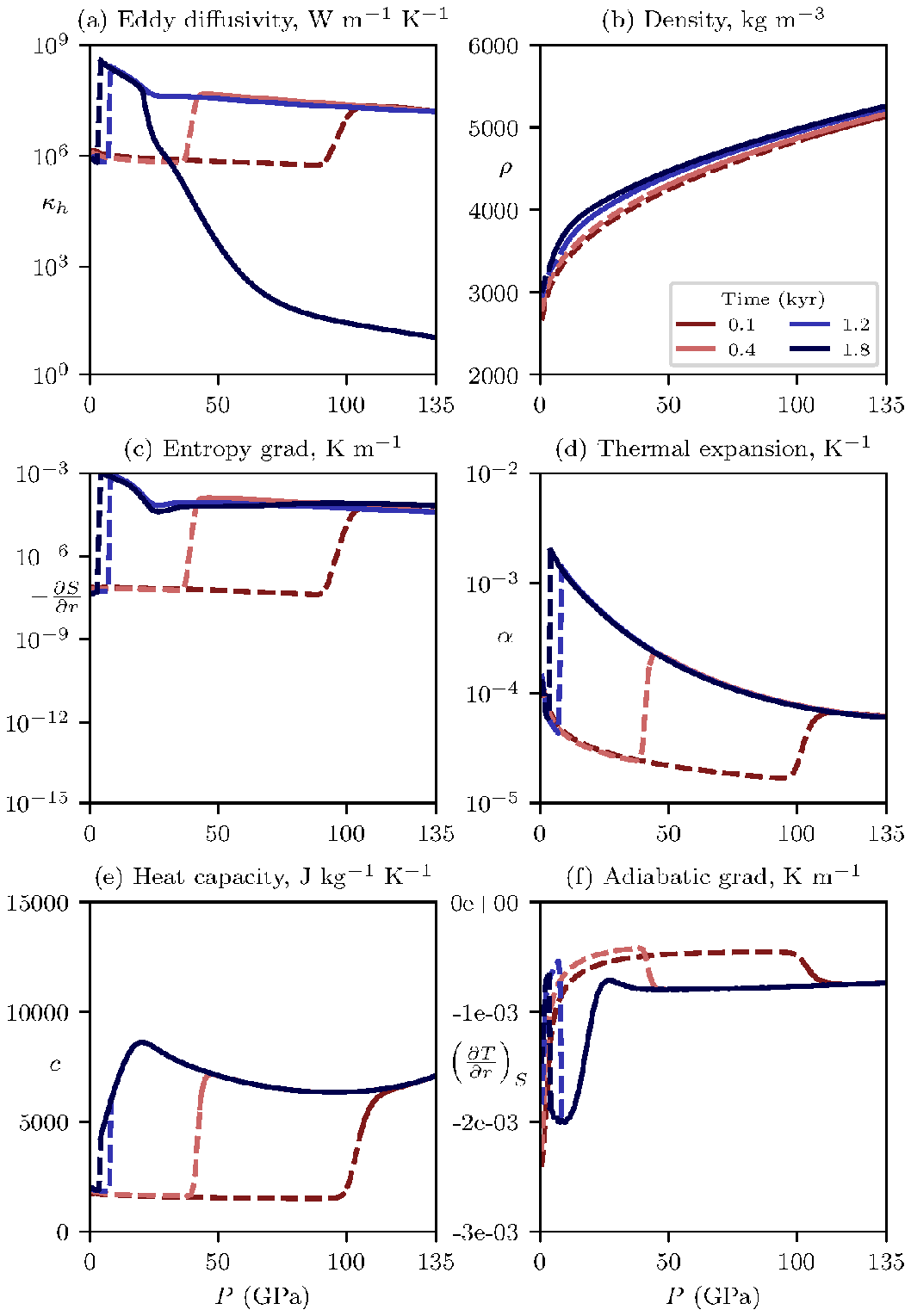}
}
\caption{Material properties for case BUM (Fig.~\ref{fig:bottom-up-modified-1}) to compare with case BU (Fig.~8).  (a) Eddy diffusivity, (b) Density, (c) Entropy gradient, (d) Thermal expansion coefficient, (e) Heat capacity, and (f) Adiabatic temperature gradient.}
\label{fig:bottom-up-modified-3}
\end{figure}

In the main manuscript we choose the mixing length to be the distance to the nearest material boundary \citep[e.g.][]{SC97}.  However, to test the sensitivity of our result to this formulation we modify case BU to use a constant mixing length whilst retaining all other parameters to be identical (case BUM).  In BUM, the mixing length is set to the 1/4 of the mantle thickness which corresponds to the average mixing length in BU.  The results for BUM are shown in Figs.~\ref{fig:bottom-up-modified-1}, \ref{fig:bottom-up-modified-2}, \ref{fig:bottom-up-modified-3} and can be directly compared with Figs.~6, 7, 8, respectively.

The results for BU and BUM are qualitatively very similar.  In fact, the evolution of BU and BUM during the earliest times before the rheological transition is reached ($\sim$ 1.6 kyr) is often visually indistinguishable (compare Fig.~6 and Fig.~\ref{fig:bottom-up-modified-1}).  During these times, the eddy diffusivity of the two cases is clearly different (Fig.~8a and Fig.~\ref{fig:bottom-up-modified-3}a) because the eddy diffusivity is a strong function of the mixing length (Eq.~8).  This affects the partitioning of the total flux between convective and mixing fluxes (Fig.~7a,c and Fig.~\ref{fig:bottom-up-modified-2}a,c) but does not strongly dictate the overall cooling behaviour.  This is because in both cases the eddy diffusivity is sufficiently large to enable efficient heat transport to the surface where the cooling rate is imposed by the ability of the planet to radiate heat.  This suggests that the high melt fraction dynamic regime of our model is not sensitive to the choice of the mixing length.

BU and BUM exhibit some differences in behaviour as the rheological transition is reached at 1.8 kyr.  In BU, the entropy is slightly higher at the base of the mantle compared to BUM (Fig.~6a and Fig.~\ref{fig:bottom-up-modified-1}a).  This is reasonable to expect given that the capacity to advect heat is partly controlled by the mixing length and the mixing length is larger at the base of the mantle in BUM than BU.  This is consistent with the total heat flux which is less in the deep mantle for BU compared to BUM (Fig.~7d and Fig.~\ref{fig:bottom-up-modified-2}d).  For BUM, a higher cooling rate in the deep mantle also enables the melt fraction to remain a monotonically increasing function with increasing radius throughout the evolution of the model.  Therefore, because the melt fraction gradient has the same sign (positive with respect to radius), the mixing flux also retains the same sign (negative) in the mixed phase region (Fig.~\ref{fig:bottom-up-modified-2}c).  In contrast, for BU the slower cooling at the base of the mantle gives rise to a melt fraction minimum around 90 GPa (Fig.~6c) and hence for pressures greater than this the mixing flux switches sign from negative to positive (Fig.~7c).  This analysis reveals that the cooling behaviour once the rheological transition is reached is more sensitive to the formulation of the mixing length than the earliest phase of rapid cooling due to liquid convection.

\end{document}